\documentclass[11pt,a4paper,twoside,titlepage]{mybooksf}

\usepackage{subfiles}
\usepackage{myreportsf}

\usepackage{graphicx}

\usepackage[dvips]{psfrag}

\usepackage[hang]{caption}
\usepackage{subcaption}

\usepackage{hyperref}

\usepackage{mathdots}
\usepackage{pifont}
\usepackage{latexsym}
\usepackage[latin1]{inputenc}
\usepackage{multicol}
\usepackage{multirow}
\usepackage{color}
\usepackage{amsmath, amssymb}
\allowdisplaybreaks[1]
\usepackage{makeidx}
\usepackage{amsfonts}
\usepackage{amsbsy}
\usepackage{epigraph}
\usepackage{xtab}
\usepackage{rotating}
\usepackage{algpseudocode}
\usepackage[withpage]{acronym}
\usepackage{algorithm}
\usepackage{eurosym}
\usepackage{longtable}
\usepackage{enumitem}
\definecolor{blackcolor}{rgb}{0,0,0}
\definecolor{darkblue}{rgb}{0,0.19,0.69}
\definecolor{gruen}{rgb}{0.16,0.69,0.38}
\usepackage{color}

\usepackage{bm}

\usepackage{nomencl}

\usepackage{rotating}
\usepackage{hhline}
\usepackage[para]{threeparttable}

\usepackage{steinmetz} 

\usepackage{tikz}
\usetikzlibrary{plotmarks}
\usetikzlibrary{backgrounds,calc, arrows, intersections}

\usetikzlibrary{shapes.geometric, arrows}
\usetikzlibrary{mindmap}

\tikzstyle{bad} = [rectangle, rounded corners, minimum width=3cm, minimum height=1cm,text centered, draw=black, fill=red!30]
\tikzstyle{good} = [rectangle, rounded corners, minimum width=3cm, minimum height=1cm,text centered, draw=black, fill=green!30]
\tikzstyle{process} = [rectangle, minimum width=4cm, minimum height=1cm, text centered, text width=4cm, draw=black,fill=gray!30]
\tikzstyle{arrow} = [thick,->,>=stealth]
\tikzstyle{every node}=[font=\small]

  
\usepackage[disable]{todonotes}

\renewcommand{\Re}{\operatorname{Re}}
\renewcommand{\Im}{\operatorname{Im}}

\begin{document}

\frontmatter
\title{\Huge {\bf \textsf{Optimization in \\ Modern Power Systems}} \\ \vspace{0.5cm} DTU Course 31765 \vspace{0.5cm} \\ \huge{Lecture Notes} \vspace{0.5cm}}

\author{{\centering \Large Spyros Chatzivasileiadis}} 

\date{ \Large Technical University of Denmark (DTU) \vfill September \the\year \\ \vskip 20mm {\centering\includegraphics[height=1.2cm]{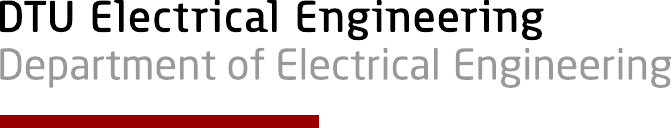} \hfill \includegraphics[height=2cm]{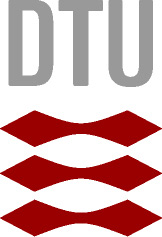}}}

\maketitle
\newpage
\thispagestyle{empty}

\chapter*{Preface}

These lecture notes aim to supplement the study material for the course ``31765: Optimization in modern power systems'' at the Technical University of Denmark (DTU). They do \emph{not} substitute the lecture slides and the discussions carried out during class. The course content is defined by the material taught in the class, which these notes aim to support to a certain extent.

The first edition of the present lecture notes was prepared for the academic year 2018-2019. Note that the material presented in these notes is a constant work in progress.

For any comments, errors, or omissions, you are welcome to contact me at \href{mailto:spchatz@elektro.dtu.dk}{spchatz@elektro.dtu.dk}.

Special thanks to the students of the 31765 course for their remarks and suggestions to improve these lecture notes.

\vspace{0.4cm}

\hfill Spyros Chatzivasileiadis

\hfill September 2018

\hfill Copenhagen, Denmark

\addcontentsline{toc}{section}{Preface}

\tableofcontents

\mainmatter


\chapter{Introduction}
\label{chap:intro}

\section{Terms and Definitions}
In this section, we define a set of terms that will be recurring across several chapters of these notes.

\subsection{Marginal cost of a generator}
Marginal cost is the cost of a generator for producing one unit of energy. The cost can be expressed in any monetary unit, e.g. Eur, USD, DKK, CHF, etc. Energy is usually expressed in MWh or kWh, depending on the size of the generator and our system. For example in a wholesale market of a transmission system, we will usually express the marginal cost as Eur/MWh (or USD/MWh, etc.). In a microgrid, we would probably go for Eur/kWh (or DKK/kWh, etc.).

\paragraph*{Limitations and Clarifications}
\begin{itemize}
  \item In the optimal power flow problem, the generator costs are assumed either linear: $f(P_G) = c_G P_G$, or  quadratic: $f(P_G) = a_G P_G^2 + b_G P_G$. The constant terms in both functions are usually neglected in the optimization, as they do not change the result of the optimization (it is just a constant offset of the final result). But: if they exist, they must be included for the calculation of the final cost.
  \item The marginal generator cost is the derivative of the generator cost curve, i.e. $\dfrac{d \, f(P_G)}{d P_G}$. As a result, if the generator cost is linear, then the marginal cost $c_G$ is a constant. If the generator cost is quadratic, the marginal cost $a_G P_G + b_G $ is linear.
  \item The generator cost as used in the optimal power flow problems is an \emph{approximation} of the true generator cost. The true cost curve of the generators $f(P_G)$ is (a) non-linear, and (b) it includes startup costs and shutdown costs, i.e. there are constant terms in $f(P_G)$ that are applied when the generator is turned on or off.
  \item Before the unbundling, vertically integrated monopolies usually approximated the cost of their conventional generators with a quadratic cost curve, as this was a closer approximation to the real cost of the conventional generator (e.g. coal, gas, oil).
  \item In electricity markets, the generator costs are usually linear. In a market environment, the marginal cost represents the bid of a generator in the market. It is much more straightforward, and more transparent (from the market operation point of view) to express this as a linear cost, i.e. every MWh you would like to buy from me will cost x USD. If we were expressing this as a quadratic cost, it would mean that we bid a linear marginal cost function  $a_G P_G + b_G $, which makes things much more complicated. Expressing this function with words, it would mean that the first MWh has a cost of $a_G+b_G$, but every subsequent MWh has a continuously higher cost by an additive term $a_G$. Such a cost function is difficult to be understood by traders and market operators during daily trading.
  \item Generators bid in the market their marginal cost, which as explained above, is a constant term (representing a linear cost curve). In that, they neglect their fixed costs, which usually include startup/shutdown and other costs, and have to approximate their variable costs with a linear function. For their fixed costs, they are usually compensated outside the day-ahead or intra-day market dispatch, through a unit commitment procedure which considers startup and shutdown generator costs.
\end{itemize}

\subsection{Contingency}
By definition, contingency refers to a future event or circumstance which is possible but cannot be predicted with certainty. In power system jargon, the term ``contingency'' refers to incidents that deviate from the planned operation and can affect the security of the power system. Contingencies include for example line outages, transformer outages, bus outages, load outages, and deviations of the power generation due to e.g. wind forecast errors or solar forecast errors. Contingencies are very often mentioned in the context of N-1 security criterion (see Section~\ref{def_sec:n1sec}).

\subsection{N-1 Security}
\label{def_sec:n1sec}
Most of the power systems around the world operate in a N-1 secure state. The N-1 security criterion stipulates that the power system must remain secure in the event of losing any single component of the system, i.e. assuming a system has N components, the system must remain secure even if it operates with N-1 components. Such a system is said to satisfy the N-1 security criterion and is called N-1 secure. In that context, a contingency is any component outage that leads to a system operation with N-1 components.

\subsection{Congestion}
The term congestion usually refers to lines or transformers. A congested line means a line that is loaded to the maximum of its transmission capacity, and as a result, it cannot carry additional power. Line (or for that purpose transformer) congestions have a negative impact on social welfare. Because of congestions, the cheapest generation units cannot produce all necessary power to cover the demand. Instead, more expensive units located downstream of a congestion must be dispatched to supply the missing power. This increases the total generation cost.

\subsection{Copperplate}
\label{def_sec:copperplate}
We usually refer to a ``copperplate network''. By that we mean that the network has infinite transmission capacity, which allows us to neglect all network constraints. The network constraints include the transmission line limit constraints, and the power flow constraints which model (a) how the power is distributed along the lines, and (b) the line losses. Neglecting these constraints gives a first approximation of the solution, simplifying a lot our calculations. The term ``copperplate'' (probably) comes from a ``copper plate'', i.e. assuming that on top of all our injection and load nodes we place a copper plate, so that all nodes are connected with each other and an infinite capacity for power transfer.

\section{Outline of the Lecture Notes}
\label{intro_sec:structure}
These notes are structured as follows:
\paragraph{Chapter 2: Economic Dispatch} This chapter presents a short overview of the market clearing based on power pools, defines the system marginal price, and the marginal generators.

\paragraph{Chapter 3: DC-OPF} This chapter introduces the DC Optimal Power Flow.

\paragraph{Chapter 4: AC-OPF} This chapter introduces the AC Optimal Power Flow.

\paragraph{Chapter 5: Semidefinite Programming} This chapter introduces Semidefinite Programming and the SDP-based Optimal Power Flow.

\section{Further reading material}
\begin{enumerate}
\item S.~Chatzivasileiadis (2012). \emph{Transmission investments in deregulated electricity markets}. Technical Report, ETH Zurich, EEH Power Systems Laboratory, May 2012.
\end{enumerate}

%
%



\chapter{Economic Dispatch}
\chaptermark{Economic Dispatch} \label{chap:economic_dispatch}

\section{What is economic dispatch?}
Imagine you are the operator of a power system: this can be a neighborhood microgrid, a large transmission system of a whole country, or something in between. Your job is to supply electricity to all loads in your system with the minimum possible cost. What do you do?

First, you need to know how much the electric demand from all your loads will be. Second, you need information about your generators: which of them are available, how much electric energy can they produce, and at what cost? When you have this information, your job is to decide which generators will be dispatched so that you cover all electric demand at minimum cost \footnote{This of course assumes that the total energy that can produced by your generators exceeds the total load demand at all times. In most power systems this is true. In the opposite case, the operator will have to decide not to serve some load (this is called load shedding or load curtailment)}.

Generally speaking, ``Economic Dispatch'' describes the process of selecting which generators to ``dispatch'' in order to achieve the most economic operation for your power system. This is the process that most system operators traditionally carried out, before the unbundling  of the power sector, i.e. when utilities were vertically integrated monopolies owning both the generation and the network infrastructure. After the unbundling, ``Economic Dispatch'' is the process that electricity markets operating under the ``Power Exchange'' or ``Power Pool'' use. \todo[inline]{Examples of markets using an Economic Dispatch algorithm...}

Over the years, industry and academia have used the term ``Economic Dispatch'' to describe variations of the same process, e.g. some economic dispatch algorithms might include network constraints and others might not, some algorithms may consider ramp constraints of the generators and others may not, etc. Most literature however converges to the following definition of economic dispatch, which we will also use for the purpose of these lecture notes:

\paragraph*{Economic Dispatch:} the optimization process that determines the operation of the least-cost available generators, given (a) the total electric demand, and (b) the minimum and maximum operation limits of each generator.

As you can easily see, this definition excludes the consideration of any network constraints (e.g. line limits), any additional generation constraints (e.g. ramp limits), and any additional security constraints.

Having defined our problem, we must now see how we can solve it. The economic dispatch problem can be solved both graphically and through an optimization procedure. Given that you know nothing about optimization how would you solve it?

\section{Merit-Order Curve}
\begin{figure}[!h]
  \centering
  \begin{tikzpicture}
   \draw[thick, ->] (0,0)--(9,0);
   \draw[thick, ->] (0,0)--(0,6);
   \draw (0,1)--(1,1);
   \draw (1,1)--(1,2);
   \draw[dashed] (1,0)--(1,1);
   \draw (1,2)--(4,2);
   \draw (4,2)--(4,4);
   \draw[dashed] (4,0)--(4,2);
   \draw (4,4)--(6,4);
   \draw (6,4)--(6,5);
   \draw[dashed] (6,0)--(6,4);
   \draw (6,5)--(8,5);
   \draw[dashed] (8,0)--(8,5);
   \node [below] at (0,0) {0};
   \node [below] at (1,0) {$A$};
   \node [below] at (4,0) {$B$};
   \node [below] at (6,0) {$C$};
   \node [below] at (8,0) {$D$};
   \node [left] at (0,1) {$c_{G1}$};
   \node [left] at (0,2) {$c_{G2}$};
   \node [left] at (0,4) {$c_{G3}$};
   \node [left] at (0,5) {$c_{G4}$};
   \node [left] at (0,6) {price};
   \node [below] at (9,0) {power};
   \node at (0.5,0.5) {$G_1$};
   \node at (2.5,0.5) {$G_2$};
   \node at (4.5, 0.5) {$G_3$};
   \node at (7, 0.5) {$G_4$};
   \draw[dashed,red] (5, -0.3)--(5,5.5);
   \node [below] at (5, -0.3) {\textcolor{red}{$P_D$}};
   \draw [dashed, red] (0,4)--(4,4);
  \end{tikzpicture}
  \caption{Economic dispatch based on the merit-order curve.}
  \label{ed_fig:meritordercurve}
\end{figure}
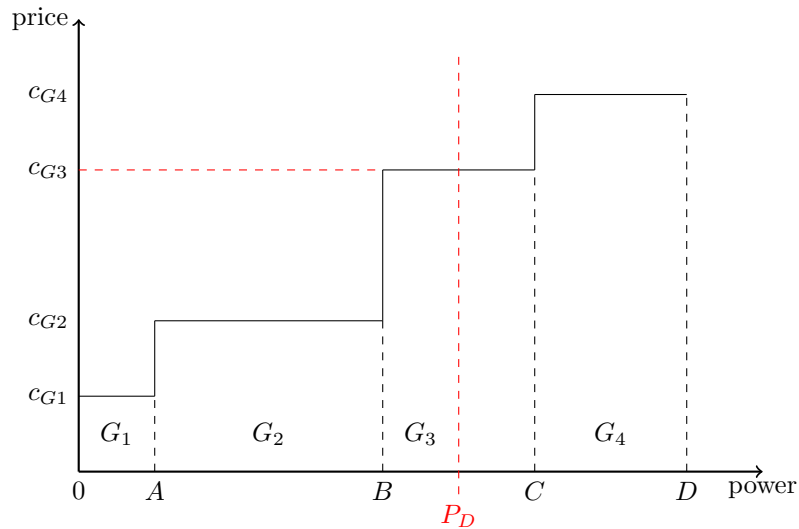

Before the unbundling of the power sector, the merit-order curve was the approach that has been traditionally used by utility engineers to decide which of their generators in the system they should dispatch to achieve the minimum operation cost. It still serves as an excellent visual tool to carry out the economic dispatch. Fig.~\ref{ed_fig:meritordercurve} shows an example of a merit-order curve.

Here are the steps to draw a merit-order curve and determine the economic dispatch. We assume linear generator costs, and, thus, constant marginal costs for each generator.
\begin{enumerate}
  \item Gather the marginal costs and the maximum generating capacity of each generator.
  \item Rank the generators from minimum to maximum marginal cost.
  \item Draw a cost curve, where the x-axis represents the power production and the y-axis represents the marginal generator cost.
  \item Start with the cheapest marginal cost and place one generator next to each other, also considering the maximum transmission capacity of each generator. Taking as example Fig.~\ref{ed_fig:meritordercurve}, it must hold $A=P_{G_1}^{max}$, $B=P_{G_1}^{max}+P_{G_2}^{max}$, $C=B+P_{G_3}^{max}$, $D=C+P_{G_4}^{max}$, and $c_{G_1} \leq c_{G_2} \leq c_{G_3} \leq c_{G_4}$.
  \item Find the intersection of the total power demand $P_D$ on the x-axis.
  \item All the generators on the left of $P_D$ must be dispatched. The rest shall not produce any power.
  \item The generator used to meet the last MWh of demand is called the ``marginal generator'' (see also Section~\ref{ed_sec:marggen}).
  \item The marginal cost of the marginal generator is called the ``system marginal price'' (see also Section~\ref{ed_sec:smp}). This is the price that all consumers must pay, and the price that all generators which were dispatched will receive (independent of their marginal cost, which by definition must lower or equal to the system marginal price).
\end{enumerate}

\paragraph*{Why is it called merit-order curve?}
Because we rank the generators based on their ``merit''. Here, generators with lower marginal costs will have a higher merit, i.e. they are ``better'' for the economic operation of the system.

\paragraph*{Limitations}
\begin{itemize}
  \item The economic dispatch assumes a copperplate network (see \ref{def_sec:copperplate}), i.e. a \emph{lossless and unrestricted flow} of electricity from point A to point B. This means that it neglects all network constraints, including transmission line limits, line congestions, and transmission losses.
  \item In such markets, system operators receive the market outcome, i.e. the dispatch of each generator determined through the economic dispatch, and run a full AC power flow, including the \mbox{N-1} security criterion. If they identify violation of operating limits, e.g. line limits or voltage limits, they carry out redispatching measures (redispatching might be part of an ancillary services market).
\end{itemize}

\section{Formulation of the optimization problem}
Equations \eqref{ed_eq:econdisp_objfun}--\eqref{ed_eq:econdisp_eq} present the formulation of the economic dispatch problem as an optimization problem:

\begin{equation}
 \min_{P_{G_i}} \sum_i c_{G_i} P_{G_i} \label{ed_eq:econdisp_objfun}
\end{equation}
subject to:
\begin{align}
 P_{G_i}^{min} \leq P_{G_i} & \leq P_{G_i}^{max} \label{ed_eq:Pminmax} \\
 \sum_i P_{G_i}             & = P_D \label{ed_eq:econdisp_eq}
\end{align}

The objective function \eqref{ed_eq:econdisp_objfun} minimizes the total power generation cost, where $c_i$ is the marginal cost of every generator and $P_{G_i}$ is the amount of power it generates. Eq. \ref{ed_eq:Pminmax} requires that all generators must not violate their minimum or maximum limits, while \ref{ed_eq:econdisp_eq} stipulates that all generated power must be equal to the electricity demand. By the way, do you notice a discrepancy in the units of the objective function?
\newline

\fbox{
\begin{minipage}{0.9\textwidth}
In the objective function, the marginal cost is given in monetary units per units of energy, e.g. Eur/MWh. The generation output however is given in units of power, e.g. MW. This means that the resulting cost is cost per unit of time, e.g. Eur/h. In other words, the Economic Dispatch (and the Optimal Power Flow as we will see in next chapters), determine the power output of each generator for a specific time period. This time period can range from a couple of minutes, e.g. 5 minutes, one hour (usual for day-ahead markets), or up to several hours, if this is a block offer. During the specified time period, the generator is expected to provide a constant $P_G$. As you may understand, this might not be too difficult for conventional generators, which can directly control their power output, but it becomes a challenge for renewable energy sources, where their output is dependent on the weather conditions. What measures can we take, so that renewable generators can participate more easily in electricity markets?
\end{minipage}
}

\section{System Marginal Price}
\label{ed_sec:smp}
The system marginal price defines the price that all consumers connected to this system will pay. That price is the same, i.e. ``uniform'', for all consumers, and corresponds to the marginal cost of the marginal generator. As we will see in Section~\ref{ed_sec:marggen}, the marginal generator is the plant used to meet the last MWh of demand.

\section{Marginal Generator}
\label{ed_sec:marggen}

The ``marginal generator'' is the plant used to meet the last MWh of demand.
If we assume:
       \begin{itemize}
        \item linear costs, that
        \item are different for each generator (even if very slightly), and
        \item no network constraints,
       \end{itemize} then there will be exactly one marginal generator in our system
If we have linear costs and we do consider the network constraints, then we will have more than one marginal generators in our system, if (and only if\footnote{the ``if and only if'' holds for the vast majority of cases that assume different linear costs for each generator (even if slightly different). There are isolated cases where a certain combination of grid topology, line reactances, and generator costs can make the ``and only if'' statement not true.}) there is one or more line congestions. We will revisit the marginal generators in Chapter~\ref{chap:dcopf}, where we will discuss about the DC Optimal Power Flow.


\chapter{DC Optimal Power Flow}
\chaptermark{DC-OPF} \label{chap:dcopf}

In this chapter we will introduce the DC Optimal Power Flow.

\section{Why is it called DC OPF?}
The terms ``DC OPF'' and ``DC Power Flow'' date back to the '60s and have nothing to do with High Voltage DC lines, and Direct Current in general. \todo{Check if it is really from the '60s and put the reference right.}

The original power flow equations for AC systems are non-linear equations of complex numbers, having a quadratic relationship between power and voltage, e.g. the apparent power flow on a line is given by $S_{ij} = V_i^2 y_{ij}^* - V_i V_j^*y_{ij}$, where $V_i, Vj, y_{ij}$ are all complex numbers (see Chapter~\ref{chap:acopf} for more details).
Instead, the term ``DC Power Flow'' refers to a linearized form of the power flow equations.

\smallskip

\fbox{
\begin{minipage}{0.9\textwidth}
\textbf{Attention:} DC OPF has absolutely nothing to do with DC lines or DC grids. DC Power Flow denotes the linearization of the original non-linear AC Power Flow equations. As a term, it is a ``remnant'' of the 60's, when DC transmission technology was still at its infancy. It was \emph{probably} inspired from the fact that the DC power flow, as an approximation, does not consider reactive power and does not have any sinus terms, similar to DC network equations. Please be aware that additional formulations are necessary to consider DC lines and DC grids in the optimal power flow formulation (both for the DC-OPF and the AC-OPF).
\end{minipage}
}

\section{Economic Dispatch vs. DC OPF}
The difference between the Economic Dispatch, as described in Chapter~\ref{chap:economic_dispatch}, and the DC Optimal Power Flow, is that DC-OPF considers the line flows in the network and it includes constraints for the line flow limits. On the contrary, Economic Dispatch considers a copperplate network, assuming that there are no constraints for power flowing between any two points in the network.

Giving a glimpse into the following sections~\ref{dc_sec:dcpowerflow}--\ref{dc_sec:dcopfformulation}, as we will see one of the most common ways to consider the line flows is to introduce the voltage angles $\theta$ as additional variables in our problem. And in order to associate them with the active power injections, we will need to include the nodal balance equations, as shown in \eqref{dc_eq:P_B_delta_2} in matrix form. For that, we will also need to form the so-called \emph{Bus Susceptance Matrix}. But before going into this, let's start with some basics.

\section{DC Power Flow}
\label{dc_sec:dcpowerflow}

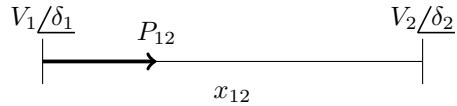
\begin{figure}[!h]
\centering
\begin{tikzpicture}
 \draw (0,-0.3) -- (0,0.3);
 \draw (0,0) -- (5,0);
 \draw (5,-0.3) -- (5,0.3);
 \draw[line width = 0.5mm, ->] (0,0) -- (1.5,0);
 \node [above] at (1.5,0.1) {$P_{12}$};
 \node [above] at (0,0.3) {$V_1 \phase{\delta_1}$};
 \node [above] at (5,0.3) {$V_2 \phase{\delta_2}$};
 \node [below] at (2.5, -0.2) {$x_{12}$};
\end{tikzpicture}
\caption{Power flow along a line (DC Power Flow)}
\label{dc_fig:lineflow}
\end{figure}

The DC power flow is an approximation of the non-linear AC power flow equations. To arrive at that approximation, we need to take a set of certain assumptions.

In a first step, if we assume that the ohmic resistance of the line is much smaller than its series reactance, then the \emph{active} power flow is given by \eqref{dc_eq:activepowerAC}. (Note: for the moment, please accept that \eqref{dc_eq:activepowerAC} holds. In Chapter~\ref{chap:acopf} we will derive the full power flow equations and show how we arrive at \eqref{dc_eq:activepowerAC}. )
\begin{align}
 P_{12} = V_1 V_2 \frac{\sin(\delta_1 - \delta_2)}{x_{12}}
 \label{dc_eq:activepowerAC}
\end{align}

Equation \eqref{dc_eq:activepowerAC} is obviously non-linear: it contains a sinus term, and a product of two variables (voltages). To linearize this equation, we make the following assumptions:

\begin{tabular}{ll}
  1) Voltage constant and at nominal value:  & $V_1 = V_2 = 1 \; \textnormal{p.u.}$ \\
  2) Angle differences are small:            & $\sin(\delta_1 - \delta_2) = \delta_1 - \delta_2$
\end{tabular}

\begin{equation}
 P_{12} = \frac{\delta_1 - \delta_2}{x_{12}} = b_{12}(\delta_1 - \delta_2)
 \label{dc_eq:Pbdelta_4}
\end{equation}

The assumptions of constant voltage and small angle differences are appropriate for lightly loaded systems, but as soon as we operate the power system closer to its limits, the voltage angle differences are no longer small. Furthermore, the assumption that the voltages remain constant traces back to the fact that in transmission systems, several nodes include infrastructure for voltage control, especially at the generator buses. This is not true for most of the nodes in distribution networks, while even in transmission networks, there is a significant number of nodes that do not have any voltage control.

Therefore, the DC power flow is a good approximation for lightly loaded systems, and possibly good enough to give a first idea of the power flows in any system, but should not be considered accurate beyond a certain operating point. Still, because of its linear characteristics (which ensures tractability for very large scale problems, guarantees convergence to global optimum, and several other favorable properties) is widely used in electricity market clearing algorithms.
\todo{maybe include some references to Pascal's and Dan's work on the accuracy of DC-OPF}

\section{Bus Susceptance Matrix}
\label{dc_sec:bussusceptancematrix}

As shown in \eqref{dc_eq:Pbdelta_4}, the active power flow is associated with the voltage angles $\delta$ and the line susceptances $b_{ij}$. In this section, we explain how we form the bus susceptance matrix, which will help us build the linear system describing how the bus voltage angles $\delta$ are associated with the bus injections $P_i$.

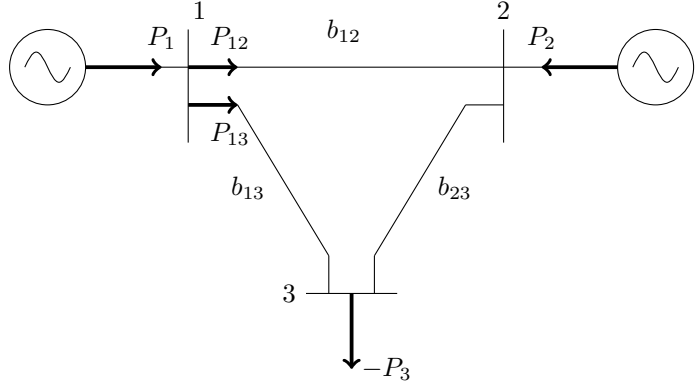
\begin{figure}[!h]
 \centering
 \begin{tikzpicture}
  \node [above] at (1,0.5) {$1$};
  \node [above] at (5,0.5) {$2$};
  \node [left] at (2.4, -3) {$3$};
  \draw (-1,0) circle (0.5cm);
  \draw (-0.5,0) -- (0.85,0);
  \draw [line width = 0.5mm, ->] (-0.5, 0) -- (0.5, 0);
  \node [above] at (0.5,0.1) {$P_1$};
  \draw (0.85,-1) -- (0.85,0.5);
  \draw (0.85,0) -- (5,0);
  \node [above] at (2.9,0.2) {$b_{12}$};
  \draw [line width = 0.5mm, ->] (0.85, 0) -- (1.5, 0);
  \node [above] at (1.4,0.1) {$P_{12}$};
  \draw (5,-1) -- (5,0.5);
  \draw (5, 0) -- (6.5,0);
  \draw [line width = 0.5mm, ->] (6.5, 0) -- (5.5, 0);
  \node [above] at (5.5,0.1) {$P_2$};
  \draw (7,0) circle (0.5cm);
  \draw (0.85,-0.5) -- (1.5,-0.5);
  \draw [line width = 0.5mm, ->] (0.85, -0.5) -- (1.5, -0.5);
  \node [below] at (1.4,-0.6) {$P_{13}$};
  \draw (1.5,-0.5) -- (2.7,-2.5);
  \node [left] at (2.0,-1.6) {$b_{13}$};
  \draw (2.7,-2.5) -- (2.7,-3);
  \draw (5,-0.5) -- (4.5,-0.5);
  \draw (4.5,-0.5) -- (3.3,-2.5);
  \node [right] at (4.0,-1.6) {$b_{23}$};
  \draw (3.3,-2.5) -- (3.3,-3);
  \draw (2.4,-3) -- (3.6,-3);
  \draw [line width=0.5mm,->] (3,-3) -- (3,-4);
  \node [right] at (3, -4) {$-P_3$};
  \draw (-1.3,0) sin (-1.15,0.2) cos (-1,0) sin (-0.85,-0.2) cos (-0.7,0);
  \draw (6.7,0) sin (6.85,0.2) cos (7,0) sin (7.15,-0.2) cos (7.3,0);
 \end{tikzpicture}
 \caption{3-bus system}
 \label{dc_fig:3bus_Bbus}
\end{figure}

From Fig.~\ref{dc_fig:3bus_Bbus}, we can derive:
\begin{align}
 P_1 & = P_{12}+P_{13}                                               \\
 P_1 & = b_{12}(\delta_1 - \delta_2) + b_{13}(\delta_1 - \delta_3)   \\
 P_1 & = (b_{12}+b_{13})\delta_1 - b_{12} \delta_2 - b_{13} \delta_3 \label{dc_eq:P_B_delta_1}
\end{align}

In matrix notation, we can express \eqref{dc_eq:P_B_delta_1} for all nodes as follows:
\begin{equation}
 {\bf P} = {\bf B} {\bf \delta},
 \label{dc_eq:P_B_delta_2}
\end{equation}

where

\begin{equation*}
 \begin{bmatrix}
  P_1 \\
  P_2 \\
  P_3
 \end{bmatrix}
 =
 \begin{bmatrix}
  b_{12}+b_{13} & -b_{12}       & -b_{13}       \\
  -b_{12}       & b_{12}+b_{23} & -b_{23}       \\
  -b_{13}       & -b_{23}       & b_{13}+b_{23} \\
 \end{bmatrix}
 \begin{bmatrix}
  \delta_1 \\
  \delta_2 \\
  \delta_3
 \end{bmatrix}
\end{equation*}

\subsection{Constructing the Bus Susceptance Matrix}
Matrix ${\bf B}$ is called the Bus Susceptance Matrix and has the following form:
\begin{equation*}
  {\bf B} =
 \begin{bmatrix}
  b_{12}+b_{13} & -b_{12}       & -b_{13}       \\
  -b_{12}       & b_{12}+b_{23} & -b_{23}       \\
  -b_{13}       & -b_{23}       & b_{13}+b_{23} \\
 \end{bmatrix}
\end{equation*}
To construct the Bus Susceptance Matrix, which is used for \emph{DC Power Flow and DC Optimal Power Flow} calculations, we can follow the following rules:
\begin{itemize}
 \item Diagonal elements $B_{ii}$: sum of all line susceptances $b_{ik}$ for all the lines connected at bus $i$. $B_{ii}=\sum_k b_{ik}$
 \item Off-diagonal elements $B_{ij}$:
       \begin{itemize}
        \item If there is a line between nodes $i$ and $j$: $B_{ij}= -b_{ij}$
        \item If there is no line between nodes $i$ and $j$: $B_{ij}= 0$
       \end{itemize}
 \item Diagonal elements are always positive
 \item Off-diagonal elements are always non-positive (zero or negative)
\end{itemize}

\section{Formulation of the optimization problem}
\label{dc_sec:dcopfformulation}
\todo{Maybe introduce the DC-OPF as a need to include flow, and then include the interdependencies between power injections and angles}
Having defined the power flow along a line, and the relationship between power injections and voltage angles through the bus susceptance matrix, we can now formulate the DC Optimal Power Flow problem.
\begin{equation}
 \min \sum_i c_i P_{G_i} \label{dc_eq:objfun}
\end{equation}
subject to:
\begin{equation}
 {\bf B} \cdot {\bm \delta}  = {\bf P_G} - {\bf P_D} \label{dc_eq:P_B_delta_3}
\end{equation}
\begin{align}
\qquad \qquad \qquad -P_{ij, max} \leq \frac{1}{x_{ij}} (\delta_i & - \delta_j) \leq P_{ij, max} &\forall i,j \in \mathcal{E} \label{dc_eq:lineflows}\\
\qquad \qquad \qquad P_{G_i}^{min} \leq   P_{G_i} \leq & P_{G_i}^{max} & \forall i \in \mathcal{N}
 \label{dc_eq:pgmax}
\end{align}

Although our objective function only minimizes the total generation cost, the DC-OPF with the standard power flow equations contains both the power generation $\bm P_G$ and the voltage angles $\bm \delta$ in the vector of the optimization variables. To accomodate that, in the objective function we add a zero cost for all angle variables.

\section{Locational Marginal Prices}
The DC-OPF is widely used in electricity markets as the market clearing algorithm. In realistic implementations it contains a large number of additional constraints and variables to account for different market products, e.g. block orders, and others. But fundamentally, all implementations build on the formulation we presented in Section~\ref{dc_sec:dcopfformulation}.

The formulation in \eqref{dc_eq:objfun} -- \eqref{dc_eq:pgmax} minimizes the total generation costs, when it receives the bids of each generator. But for this algorithm to be used by a market operator, it shall also define the price that each generator must receive. It has been found that the lagrangian multipliers of the equality constraints \eqref{dc_eq:P_B_delta_3} can be interpreted as locational marginal prices, representing the price for injecting 1 additional unit of energy at the specific node. For details on the derivation please refer to Appendix~\ref{chap:appendix_LMP}.

Locational Marginal Prices, or LMPs, are often also called ``Nodal Prices''. In these lecture notes, we will use the two terms interchangeably.

\subsection{Extraction of the Locational Marginal Prices}
Note that in order to extract the LMPs, it is \emph{required} that the objective function minimizes \emph{solely} the generation costs.

\vspace{12pt}

\fbox{
\begin{minipage}{0.9\textwidth}
If the objective function represents the minimization of the generation costs, then the lagrangrian multipliers $\nu_1, \nu_2, \ldots, \nu_N$ of the equality constraints ${\bm{P}=\bm{B\delta}}$ represent the nodal prices.
\begin{align*}
  P_1 & = B_{11}\delta_1 +B_{12}\delta_2 + \ldots +B_{1N}\delta_N \quad &:& \quad \nu_1 \\
  P_2 & = B_{21}\delta_1 +B_{22}\delta_2 + \ldots +B_{2N}\delta_N \quad &:& \quad \nu_2 \\
  &\vdots && \\
  P_N & = B_{N1}\delta_1 +B_{N2}\delta_2 + \ldots +B_{NN}\delta_N \quad &:& \quad \nu_N \\
\end{align*}
There is one lagrangian multiplier $\nu_n$ for each nodal equation, and as a result, there is one nodal price for each node.
\\
\\
The LMPs denote what is the cost for injecting one additional MWh (``marginal'') at the specific node (``locational''):
\begin{itemize}
  \item This is the price that a consumer should pay at that node
  \item This is the price that a producer should get paid at that node
\end{itemize}
\end{minipage}
}
\hfill
\vspace{12pt}
Most common solvers calculate the Lagrangian multipliers along with the optimal solution, so it is usually straightforward to extract the nodal prices while solving an optimal power flow problem.

\section{Key Discussion Points}
This section summarizes a set of key points, that are important to remember when we formulate or implement the DC-OPF.

\begin{figure}
  \centering
  \includegraphics[width=0.7\textwidth]{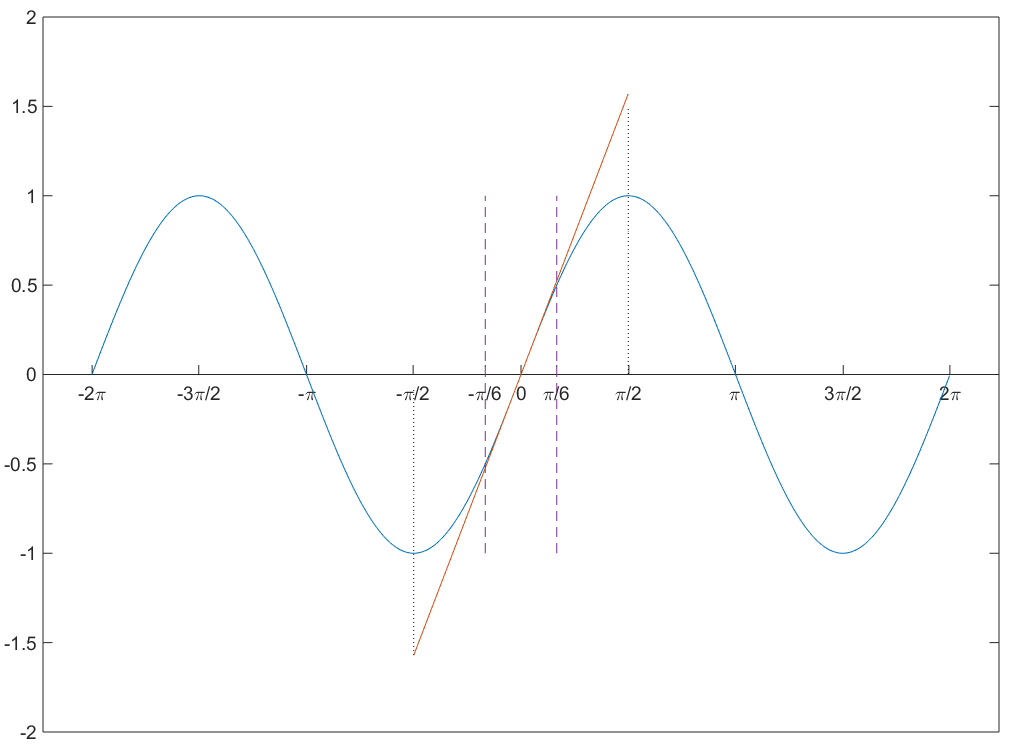}
  \caption{$y=\sin\delta$ (blue line) vs $y=\delta$ (red line). We see that the approximation $\sin \delta = \delta$ holds well for angles up to $\pi/6$. Beyond that point, the accuracy of the approximation decreases substantially.}
  \label{dc_fig:sind_equal_d}
\end{figure}

\paragraph{$\delta$ is in rad}
The approximation $\sin \delta \approx \delta$ holds for small angles $\delta$ and as long as $\delta$ is expressed in rad. Indeed, let's assume an angle $\delta=30^{\circ}$. It is $\sin 30^{\circ}=0.5$ and $\delta=\frac{\pi}{6}=0.5236 \text{ rad}$. So, as long as $\delta$ is expressed in rad, we can say that $\sin \frac{\pi}{6} \approx \frac{\pi}{6} =0.5236$.
Fig.~\ref{dc_fig:sind_equal_d} presents graphically the accuracy of the approximation $\sin \delta = \delta$. As we can observe, beyond $\delta = \frac{\pi}{6}$ the accuracy of the approximation decreases.

If you remember from \eqref{dc_eq:Pbdelta_4}, the approximation $\sin (\delta_1 - \delta_2) = \delta_1 - \delta_2$ was made for the voltage angle difference between neighboring buses. So, what is important to remember is that if two neighboring buses have a voltage angle difference beyond $\pi/6$ this approximation is no longer accurate.

Having said that, it is important to stress that both the DC power flow and the DC Optimal Power Flow are widely used in practice for highly loaded systems and voltage angle differences that go beyond $\pi/6$ (e.g. market clearing). However, it is important to remember that in order to make sure that the system is secure, and no line flow constraints or voltage constraints are violated, we need to plug the DC-OPF results in an AC power flow or an AC-OPF and check if any violations occur. If violations occur, the system operator needs to perform a redispatching.

\paragraph{If $\delta$ is in rad, then $\bm{P}$ is in per unit}
From \eqref{dc_eq:P_B_delta_3} it holds that $\bm{B} \cdot \bm{\delta} = \bm{P}$.
All elements of the Bus Susceptance Matrix ${\bf B}$ are in the per unit system (p.u.).
Since rad is dimensionless, if $\delta$ is in rad, all elements of vectors ${\bf P}$ must be in p.u.

Note: It is possible to use kW or MW for ${\bf P}$ (instead of p.u.), but in that case $\delta$ will not represent voltage angles. It is generally advised to keep ${\bf P}$ in p.u. (and as a result $\delta$ in rad). There are two main reasons for that, both related to the numerical accuracy of the optimization solvers, as we explain in the following paragraph.

\paragraph{Scaling, numerical accuracy, and the per unit system}
Scaling of variables and constraints is important for all numerical solvers, including the solvers used for the solution of optimization problems. In general, we try to keep variables and constraints to values close to [-10,\ldots,10], ideally in the range [-1,\ldots,1]. There are two main reasons for that:
\begin{itemize}
  \item First, and most importantly, the number of floating point numbers that can be represented by computers within the range $[-1,1]$ is almost as large as anywhere else in $\mathcal{R}$, i.e. $(-\infty,-1)$ and $(1,\infty)$. By having floating point values within the region $[-1,1]$, the accuracy of the numerical solver increases substantially. The per unit system is helpful for that: it transforms large values of 100's of MW to values around 1 p.u.
  \item The numerical solvers converge to a solution after several iterations. In every iteration they check the new solution against the solution of the previous iteration. If the difference is smaller than a tolerance value then they stop. Tolerances are used both for optimality (objective function) and feasibility (constraints).
  Default values are around $10^{-6}$.  Imagine now that you express the active power in Watts, and $P_1=100$ MW.
  Then while evaluating the constraint $B_{11}\delta_1+B_{12}\delta_2+B_{13}\delta_3 = P_1$ the solver will try to achieve an accuracy of 100'000'000.0000001 W.
  If you have expressed $P_1$ in p.u with baseMVA = 100 MW, then the solver will try to achieve an accuracy of 1.0000001 p.u. which is equal to 100'000'010 W. When we are dealing with Megawatts, the difference of some Watts is almost negligible. By appropriate scaling (and the per unit system is a helpful scaling trick in this case) we can develop optimization problems that solve much more efficiently.
\end{itemize}

Scaling is so important that almost all optimization solvers do some kind of internal scaling, in order to bring the problem in an appropriate form to solve efficiently. So, essentially, even if our problem is badly scaled, state-of-the-art solvers will try to bring it to an appropriate form. Still, we know much better than any generic solver the problem we want to solve. As a result, it is usually much more effective if we try to do some scaling of the problem ourselves than rely only on the solver.

For example, if our objective function $\min \sum_i c_i P_{G_i}$ evaluates to hundreds of thousands of dollars per hour, we can add a scaling factor, e.g. $\min \frac{1}{1000}\sum_i c_i P_{G_i}$. The result of the optimization, i.e. the values of each variable will be exactly the same, as the feasible space depends on the constraints. We only need to remember that as soon as we have obtained the solution, we need to multiply the objective function value by 1'000 to find the actual cost.

\paragraph{${\bf B}$ in DC-OPF vs ${\bf Y}$ in AC-OPF}

We need to be careful when we are building the Bus Susceptance matrix ${\bf B}$. The elements of ${\bf B}$ are based on the inverse of the reactances, i.e. $B_{ij} = -b_{ij}$ (for $i \neq j$) and $b_{ij} = \dfrac{1}{x_{ij}}$. When calculating $b_{ij}$ we must not assume that it is the imaginary part of the admittance $y_{ij}=g_{ij}+jb_{ij}$.
The term $y_{ij}$ comes from inverting the complex impedance $z_{ij}=r_{ij}+jx_{ij}$, and in that case $\Im\{y_{ij}\}$ will be negative.

Example: Assume $z_{ij}=r_{ij}+jx_{ij}=0.02+j0.4$ p.u. Then:
\begin{itemize}
  \item $y_{ij}=\dfrac{1}{z_{ij}} = 10-j20$
  \item but, for the purposes of the DC-OPF (and DC Power Flow)$b_{ij} = \frac{1}{0.4} = 20$, which is different from the $\Im\{y_{ij}\}$. In Chapter~\ref{chap:acopf} we will see where this difference comes from.
\end{itemize}

\section{Power Transfer Distribution Factors (PTDF)}

As we mentioned earlier in this chapter, the main difference between the Economic Dispatch algorithm, and the DC-OPF is that the DC-OPF considers how the power is distributed along the lines, and includes all line flow constraints (using a linear approximation of the active power flow). To do that, we needed to intoduce additional variables, the voltage angles $\delta$, and additional constraints on top of the line flow constraints, i.e. the nodal balance equations $\bm{P}=\bm{B\delta}$.

But can we calculate the line flow constraints without the need to compute $\bm{\delta}$? Yes! By reformulating our constraints and using the Power Transfer Distribution Factors (PTDF).

The PTDF is a linear sensitivity that represents the marginal change of the active power flow on a line if we apply a marginal increase of the power injection at a node. Take for example the 3-bus system of Fig.~\ref{dc_fig:3bus_Bbus}. Assume that the slack bus is Bus 1. Then $PTDF_{13,2}=0.33$ means that if we inject 1~MW at Bus 2 (and we receive it at Bus 1 -- or, in other words, Bus 1 reduces its power injection by 1 MW), then the power flow along line 1-3 will increase by 0.33~MW.  For every tuple $\langle \text{line, node} \rangle$ we have a different power transfer distribution factor (PTDF).
\todo{Add an example about the calculation of PTDF. Tasks: a) Calculate the $PTDF_{ij,m}$, b) calculate the $PTDF_{ij,mn}$, c)change the slack bus and recalculate, d) calculate the PTDF based on the matrix inversion}

In more formal terms, the change in the flow of line $ij$ associated with a power injection at node $m$ and an equivalent withdrawal at the slack bus is given by:
\begin{equation}
 \Delta P_{ij} = PTDF_{ij,m} \Delta P_{m}
 \label{dc_eq:PTDF_1}
\end{equation}

Imagine now that we want to calculate what is the change of the flow in line $ij$, if we inject 10~MW at node $m$ (e.g. a generator node) and receive it at a random node $n$ (e.g. a load node), other than the slack bus. Since we are working within the DC-OPF context, we have linear equations and we neglect transmission losses. As a result, $\Delta P_{m \rightarrow n}= 10 MW$ is the same as injecting in bus $m$ 10~MW and receiving them at slack bus $k$, and then injecting at slack bus $k$ 10 MW and receiving them at bus $n$. This can be written as: $\Delta P_m = 10 \text{ MW}$, $\Delta P_k = -10 + 10 = 0 \text{ MW}$, $\Delta P_n = -10 \text{ MW}$, where the positive sign (+) means power injection and the negative (-) withdrawal.

Following \eqref{dc_eq:PTDF_1}, we can write:
\begin{align}
  \Delta P_{ij} & = PTDF_{ij,m} \Delta P_{m} + PTDF_{ij,n} \Delta P_{n} \notag \\
                & = PTDF_{ij,m} \cdot 10 + PTDF_{ij,n} \cdot (-10) \notag \\
                & = (PTDF_{ij,m} - PTDF_{ij,n}) \cdot 10 \notag \\
                & = PTDF_{ij,mn} \cdot 10 \Rightarrow \notag \\
& \notag\\
  \Delta P_{ij} & = PTDF_{ij,mn} \cdot \Delta P_{m \rightarrow n}, \label{dc_eq:PTDF_2} \\
                & \qquad \text{  where  } PTDF_{ij,mn} = PTDF_{ij,m}-PTDF_{ij,n} \notag
\end{align}

From \eqref{dc_eq:PTDF_2}, it becomes obvious that the calculation of the line flows are independent of the choice of the slack bus. To be more specific, although the values of $PTDF_{ij,m}, PTDF_{ij,n}$ will change by selecting a different slack bus, the difference $PTDF_{ij,mn} = PTDF_{ij,m}-PTDF_{ij,n}$ will remain the same irrespective of the choice of slack bus.

In case now we want to calculate the total flow over a line, from \eqref{dc_eq:PTDF_1} it follows that:
\begin{equation}
 P_{ij}=\sum_{m} PTDF_{ij,m} P_m, \label{dc_eq:PTDFlineflow}
\end{equation}
where $P_m$ are the injections/withdrawals at every bus of the system.

\subsection{Calculation of the PTDF}
The PTDF of line $i-j$ for power injection at node $m$ and power withdrawal at node $n$ is given by:
 \begin{equation}
  PTDF_{ij,mn} = \frac{X_{im}-X_{jm}-X_{in}+X_{jn}}{x_{ij}}
  \label{dc_eq:PTDF_3}
 \end{equation}
 where
 \begin{tabular}{ll}
  $x_{ij}$ & reactance of the transmission line conneting node $i$ and node $j$ \\
  $X_{im}$ & is the entry in the $i$th row and the $m$th column of the bus \\ & reactance matrix $\mathbf{X_{bus}}$\\
 \end{tabular}

\paragraph{Bus Reactance Matrix}
To calculate \eqref{dc_eq:PTDF_3}, we need first to calculate the bus reactance matrix. The Bus Reactance Matrix is essentially the inverse of the Bus Susceptance Matrix, as introduced in Section~\ref{dc_sec:bussusceptancematrix}. However, the Bus Susceptance Matrix $B$ is a singular matrix, and thus it is not invertible. In order to invert it and obtain the bus reactance matrix, we follow the procedure below:
\begin{enumerate}
  \item From the $N_{bus} \times N_{bus}$ size Bus Susceptance Matrix ${\bf B_{bus}}$ remove the row and column that correspond to the slack bus to form $\bm{\tilde{B}_{bus}}$ with size $(N_{bus}-1) \times (N_{bus}-1)$
  \item Invert $\bm{\tilde{B}_{bus}}$, which still has size $(N_{bus}-1) \times (N_{bus}-1)$
  \item Add a row and column of zeros at the row and column corresponding to the slack bus to form the $N_{bus} \times N_{bus}$ size matrix  $\bm{X_{bus}} = \bm{\tilde{B}_{bus}^{-1}}$
\end{enumerate}
Having obtained the bus reactance matrix, we can now calculate each signle PTDF element through \eqref{dc_eq:PTDF_3}.

If, however, we want to calculate the whole PTDF matrix, i.e. for all lines and nodes, there is a simpler procedure. As shown in \eqref{dc_eq:PTDF_4}, the PTDF matrix is the product of the Bus Reactance Matrix and the Line Susceptance Matrix:
\begin{equation}
  {\bf PTDF} = {\bf B_{line}} {\bf \tilde{B}_{bus}}^{-1} \label{dc_eq:PTDF_4}
\end{equation}

We have outlined above how to calculate the Bus Reactance Matrix. In the paragraph below, we will focus on the Line Susceptance Matrix.

\paragraph{Line Susceptance Matrix}
The line susceptance matrix collects all linear equations (as shown in \eqref{dc_eq:lineflows}) that relate the voltage angles $\delta_i$ with the line flows $P_{\text{line},ij}$. In other words, the Line Susceptance Matrix helps us write in a compact form the linear equations for all line flows as follows:
\begin{equation}
  \mathbf{P}_{\text{line}}=\mathbf{B}_{\text{line}}\bm{\delta} \Leftrightarrow P_{\text{line},ij} = \frac{1}{x_{ij}}(\delta_i-\delta_j) \quad \forall i,j \in \mathcal{E}
  \label{dc_eq:Bline_lineflow}
\end{equation}

How do we form the line susceptance matrix?
\begin{itemize}
  \item $\mathbf{B}_{\text{line}}$ has dimensions $L \times N$, where $L$ is the number of lines and $N$ the number of nodes.
  \item every row has only two non-zero elements; these are at the columns corresponding to the nodes each line connects
\end{itemize}

\paragraph{Example}
What is the Line Susceptance Matrix for the 3-bus system in Fig.~\ref{dc_fig:3bus_Bbus}?

\paragraph{Solution}
\begin{equation}
\bm{B}_{\text{line}} =
\begin{bmatrix}
b_{12} & -b_{12} & 0 \\
b_{13} & 0 & -b_{13}\\
0 & b_{23} & -b_{23} \\
\end{bmatrix}
\notag
\end{equation}
Attention! The Line Susceptance Matrix in this example happens to be a $3 \times 3$ matrix (i.e. square), because our example has 3 lines and 3 buses. Contrary to the Bus Susceptance Matrix which is always square, however, the Line Susceptance Matrix is most often \emph{not} square. Most transmission systems are meshed power systems and have more lines than buses. This results to thin matrices (i.e. more rows than columns) for the Line Susceptance Matrix.

\paragraph{Forming the Line Flow Constraints with PTDF\\}
 From \eqref{dc_eq:P_B_delta_3}, it is:
 \begin{equation}
  \bm{\delta} = {\bf \tilde{B}_{bus}}^{-1} ({\bf P_G - P_D})
  \label{PTDFcalc1}
 \end{equation}
 while from \eqref{dc_eq:Bline_lineflow}, we can form the line flow constraints as follows:
 \begin{equation}
  {\bf B_{line}} \bm{\delta} \leq {\bf P_{line}^{max}}
  \label{PTDFcalc2}
 \end{equation}
 Then, replacing $\bm{\delta}$ in \eqref{PTDFcalc2} with \eqref{PTDFcalc1}, we define:
 \begin{equation}
  {\bf PTDF} = {\bf B_{line}} {\bf \tilde{B}_{bus}}^{-1}
 \end{equation}
 and we get:
 \begin{equation}
  {\bf PTDF} \left({\bf P_G - P_D}\right) \leq {\bf P_{line}^{max}} \label{dc_eq:PTDFlineflow_matrix}
 \end{equation}
In Section~\ref{dc_sec:dcopf_ptdf}, we will see how we formulate a DC-OPF program using the PTDFs and equation \eqref{dc_eq:PTDFlineflow_matrix} as a constraint.

\subsection{Example}

\begin{figure}[H]
     \centering
   \includegraphics[width=0.7\columnwidth]{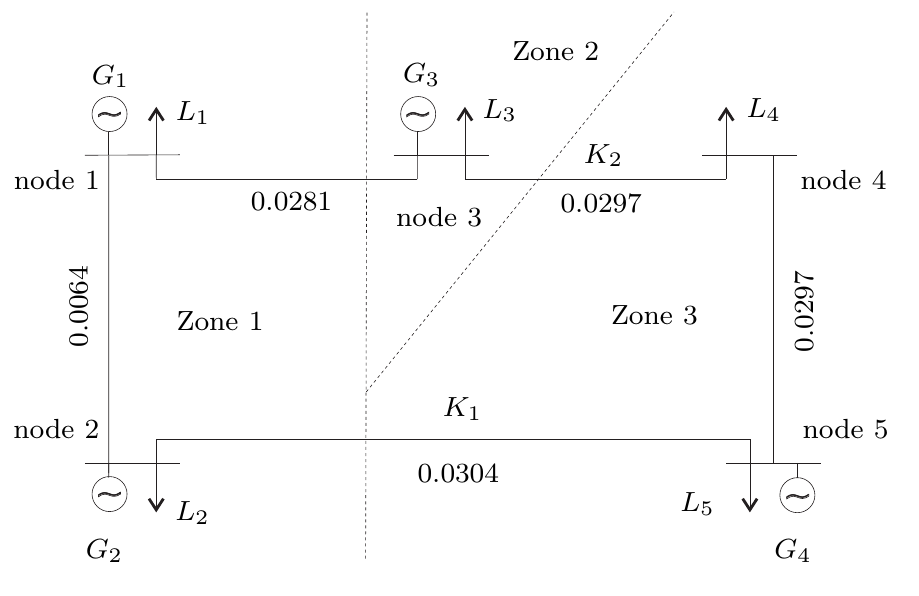}
\caption{5-bus system. The numerical values correspond to the line reactances expressed in p.u. \tiny{Figure taken from: T.Krause, \emph{Evaluating Congestion Management Schemes in Liberalized Electricity Markets Applying Agent-based Computational Economics}, PhD Thesis, ETH Zurich, 2007.}}
   \label{dc_fig:PTDF_example}
 \end{figure}

Calculate the PTDF matrix of the 5-bus system shown in Fig.~\ref{dc_fig:PTDF_example}.

\paragraph{Solution\\}

\begin{tabular}{r|rrrrr}
     & 1 & 2 & 3 & 4 & 5 \\
     \hline
1-2  & 0 & -0.9432  & -0.2496  & -0.5133  & -0.6732  \\
1-3  & 0 & -0.0568  & -0.7504  & -0.4867  & -0.3268  \\
2-5  & 0 &  0.0568  & -0.2496  & -0.5133  & -0.6732  \\
3-4  & 0 & -0.0568  &  0.2496  & -0.4867  & -0.3268  \\
4-5  & 0 & -0.0568  &  0.2496  &  0.5133  & -0.3268  \\
\end{tabular}


\section{DC-OPF formulation based on PTDF}
\label{dc_sec:dcopf_ptdf}
Based on the above, we can calculate all line flows with the help of \eqref{dc_eq:PTDFlineflow_matrix}, without the need of the nodal balance equations \eqref{dc_eq:P_B_delta_3} or the additional variables $\delta$. The DC-OPF based on PTDFs is formulated as follows:

\begin{equation}
 \min \sum_i c_i P_{G_i} \notag
\end{equation}
subject to:
\begin{align}
 P_{G_i}^{min} \leq  & P_{G_i} \leq P_{G_i}^{max} \\
 \sum_i P_{G_i} - & \sum_i P_{D_i} = 0 \label{dc_eq:loadgen_balance} \\
 -{\bf P_{line}^{max}} \leq {\bf PTDF} & ({\bf P_G} - {\bf P_D}) \leq {\bf P_{line}^{max}}
\end{align}

Note that we still need to have a single equation which will ensure that the total generation equal the total demand, as shown \eqref{dc_eq:loadgen_balance}. But we do not have a separate equality constraint for each node; this is not necessary.

The DC-OPF based on PTDF formulation is currently used for the flow-based market coupling of the European markets. Each node corresponds to a zone (usually a country) and PTDFs are derived for the interconnections between countries.


\chapter{AC Optimal Power Flow}
\chaptermark{AC-OPF} \label{chap:acopf}

\section{What is AC Optimal Power Flow?}
The AC Optimal Power Flow is an optimization algorithm that considers the full AC power flow equations. Assuming that the model parameters are correct, this is the most accurate representation of the power flows in a system. This means that the setpoints determined by the optimization correspond as close as possible to reality. The US
Federal Energy Regulatory Commission (FERC) states that the ultimate goal for an electricity market software should be a security-constrained AC optimal power flow with unit commitmnent \cite{FERC_acopf}. FERC goes on further to indicate that a good optimization solution technique could potentially save tens of billions of dollars annually \cite{FERC_acopf}.

In this chapter we will introduce the basic formulation of the AC optimal power flow. This is the foundation of any AC optimal power flow algorithm (e.g. security-constrained AC-OPF, unit commitment AC-OPF, optimal AC transmission switching, etc.)

Compared to the DC-OPF formulation that we discussed in the previous chapter, the benefits of the AC-OPF are (i) increased accuracy, (ii) considers voltage, (iii) considers reactive power, (iv) considers currents, (v) considers transmission losses (and other types of losses). However, there are also drawbacks. The AC power flow equations are quadratic equations (since the power is dependent on the square of the voltage), and if we include them in an optimization problem as equality constraints, they result to a non-linear non-convex problem. Non-convex problems are in general much harder to solve, and there is no guarantee that the solver can find the global minimum. In the past years, serious efforts have been made to ``convexify'' the AC-OPF. The usual procedure to do that is to relax the non-convex problem to a convex one (i.e. by defining a convex function around the non-convex function), then solve the convex problem, and obtain the global optimum. If the global optimum is feasible for the original problem, then we have identified the optimal solution. If the global optimum lies outside the feasible space, then we employ a series of different techniques to recover a feasible solution as close as possible to the global optimum. In the next chapter, we will discuss the convex relaxations of the AC-OPF, focusing on the Semidefinite Programming formulation.

One of the central elements for the implementation of the AC-OPF is the modeling of the transmission lines. In the following sections we outline the modeling of the transmission lines, and the derivation of the AC power flow equations. For more details, the interested reader is encouraged to refer to dedicated textbooks in the field (e.g. \cite{GloverOverbyeSarma}).

\section{AC Power Flow}
\label{ac_sec:acpowerflow}
\subsection{Modeling transmission lines: the $\pi$-model}
The most usual representation of a transmission line in power systems is the so-called $\pi$-model. For lines between 25~km and 250~km, the  $\pi$-model is the most prefered modeling approach. Although there are both simpler and more complex modelling approaches for the transmission lines (see \cite{GloverOverbyeSarma}), in this chapter we will focus only on the $\pi$-model for transmission lines, as this is the model used by the vast majority of AC-OPF software.

\begin{figure}[!h]
  \centering
  \begin{tikzpicture}
   \draw (0,-0.3) -- (0,0.3);
   \draw (0,0) -- (6,0);
   \draw (6,-0.3) -- (6,0.3);
   \draw (1,0) -- (1,-2.5);
   \draw (5,0) -- (5,-2.5);
   \draw (0.8,-2.5) -- (1.2,-2.5);
   \draw (0.85,-2.55) -- (1.15,-2.55);
   \draw (0.9,-2.6) -- (1.1,-2.6);
   \draw (5,0) -- (5,-2.5);
   \draw (4.8,-2.5) -- (5.2,-2.5);
   \draw (4.85,-2.55) -- (5.15,-2.55);
   \draw (4.9,-2.6) -- (5.1,-2.6);
   \node [above] at (0,0.3) {$\cmp{V}_i$};
   \node [above] at (6,0.3) {$\cmp{V}_j$};
   \node [below] at (3, -0.1) {$R_{ij}+jX_{ij}$};
   \node [right] at (1, -1.5) {$\dfrac{jB_{ij}}{2}$};
   \node [left] at (5, -1.5) {$\dfrac{jB_{ij}}{2}$};
  \end{tikzpicture}
 \caption{$\pi$-model of the line}
 \label{ac_fig:pimodel}
\end{figure}
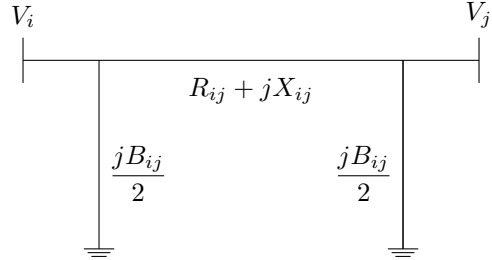

Assume a transmission line connected between the nodes $i$ and $j$ in a network, as shown in Fig.~\ref{ac_fig:pimodel}. The $\pi$-model of a transmission line consists of a series impedance $R_{ij}+jX_{ij}$ connected at the nodes $i$ and $j$, and two equal shunt susceptances, one connected at node $i$ and the other at node $j$.

We usually represent the $\pi$-model by the series admittance: $\cmp{y}_{ij} = \dfrac{1}{R_{ij}+jX_{ij}}$, and by the shunt susceptances at both ends of the line:  $\cmp{y}_{sh,i} = \cmp{y}_{sh,j} = j\dfrac{B_{ij}}{2}$.

\subsection{Current flow along a line}

Assume a current entering node $i$ in Fig.~\ref{ac_fig:pimodel}. How will it flow along the line? A large part of it will flow along the series admittance $y_{ij}$, but there will also be some current flowing along the shunt susceptance $y_{sh,i}$. The total current entering node $i$ will be equal to:

\begin{equation}
I_{i\rightarrow j}  = I_{sh,i}+I_{ij} = \cmp{y}_{sh,i}\cmp{V}_i+\cmp{y}_{ij}(\cmp{V}_i-\cmp{V}_j)
\label{ac_eq:current_ij}
\end{equation}

In matrix form \eqref{ac_eq:current_ij} is written as:
\begin{equation}
  I_{i\rightarrow j} = \begin{bmatrix}
  \cmp{y}_{sh,i}+\cmp{y}_{ij} & -\cmp{y}_{ij}
  \end{bmatrix}
  \begin{bmatrix}
  \cmp{V}_i \\
  \cmp{V}_j
  \end{bmatrix}
  \label{ac_eq:current_ij_matrix}
\end{equation}

Similar is the case for a current entering the line at node $j$. In that case the total current will be equal to:
\begin{equation}
I_{j\rightarrow i} = I_{sh,j}+I_{ji} = \cmp{y}_{sh,j}\cmp{V}_j+\cmp{y}_{ij}(\cmp{V}_j-\cmp{V}_i)
\label{ac_eq:current_ji}
\end{equation}

Again, in matrix form \eqref{ac_eq:current_ji} is written as:
\begin{equation}
I_{j\rightarrow i} = \begin{bmatrix}
 -\cmp{y}_{ij} & \cmp{y}_{sh,j}+\cmp{y}_{ij}
 \end{bmatrix}
 \begin{bmatrix}
 \cmp{V}_i \\
 \cmp{V}_j
 \end{bmatrix}
 \label{ac_eq:current_ji_matrix}
\end{equation}

Comparing \eqref{ac_eq:current_ij} with \eqref{ac_eq:current_ji} (or, equivalently, \eqref{ac_eq:current_ij_matrix} with \eqref{ac_eq:current_ji_matrix}), we see that $ I_{i\rightarrow j} \neq -I_{j\rightarrow i}$. This means that part of the current that is injected at node $i$ never arrives at node $j$, and vice versa. The difference between $ I_{i\rightarrow j}$ and $-I_{j\rightarrow i}$ is part of the losses along the transmission line.



%



\subsection{Line Admittance Matrix}

Looking at \eqref{ac_eq:current_ij_matrix} and \eqref{ac_eq:current_ji_matrix}, the question soon arises if it is possible to organize all current flows in a vector, and form a set of linear equations in a matrix form. Indeed, similar to the line susceptance matrix we saw in chapter \ref{chap:dcopf} we can form the line admittance matrix.

As shown in \eqref{ac_eq:LineAdmitMatrix_compact}, the line admittance matrix $\mathbf{Y}_{\text{line}}$ links the bus voltages $V_1, \ldots, V_n$ to the current flows $I_{1\rightarrow 2}, \ldots, I_{m\rightarrow n}$:

\begin{equation}
  \mathbf{I}_{\text{line}} = \mathbf{Y}_{\text{line}} \mathbf{V}
  \label{ac_eq:LineAdmitMatrix_compact}
\end{equation}

As already mentioned, taking a look at \eqref{ac_eq:current_ij_matrix} and \eqref{ac_eq:current_ji_matrix}, we realize that the current flowing in the direction $i \rightarrow j$ is different from the current flowing in the opposite direction $j \rightarrow i$, with the difference between the two currents being related to the losses along the line. Because of this difference, we must formulate two line admittance matrices, one for the direction $i \rightarrow j$ and one for the direction $j \rightarrow i$.

\begin{equation}
\begin{bmatrix}
  I_{1 \rightarrow 2} \\
  \vdots \\
  I_{i \rightarrow j} \\
  \vdots \\
  I_{m \rightarrow n} \\
\end{bmatrix}
=
\begin{bmatrix}
  \cmp{y}_{sh,1}+\cmp{y}_{12} & -\cmp{y}_{12} & 0 & \ldots & \ldots 0 \\
  \vdots & \vdots & \vdots & \vdots & \vdots  \\
  0\ldots & \cmp{y}_{sh,i}+\cmp{y}_{ij} & \ldots & -\cmp{y}_{ij} & \ldots 0 \\
  \vdots & \vdots & \vdots & \vdots & \vdots \\
  0\ldots & \ldots & \cmp{y}_{sh,m}+\cmp{y}_{mn} & \ldots & -\cmp{y}_{mn} \\
\end{bmatrix}
\begin{bmatrix}
  V_{1} \\
  V_2 \\
  \vdots \\
  V_{i} \\
  \vdots \\
  V_{j} \\
  \vdots \\
  V_{n} \\
\end{bmatrix}
\label{ac_eq:LineAdmitMatrix_ij}
\end{equation}

\begin{equation}
\begin{bmatrix}
  I_{2 \rightarrow 1} \\
  \vdots \\
  I_{j \rightarrow i} \\
  \vdots \\
  I_{n \rightarrow m} \\
\end{bmatrix}
=
\begin{bmatrix}
   -\cmp{y}_{12} & \cmp{y}_{sh,1}+\cmp{y}_{12} & 0 & \ldots & \ldots 0 \\
  \vdots & \vdots & \vdots & \vdots & \vdots  \\
  0\ldots & -\cmp{y}_{ij} & \ldots & \cmp{y}_{sh,i}+\cmp{y}_{ij} & \ldots0 \\
  \vdots & \vdots & \vdots & \vdots & \vdots \\
  0\ldots & \ldots & -\cmp{y}_{mn} & \ldots & \cmp{y}_{sh,m}+\cmp{y}_{mn} \\
\end{bmatrix}
\begin{bmatrix}
  V_{1} \\
  V_2 \\
  \vdots \\
  V_{i} \\
  \vdots \\
  V_{j} \\
  \vdots \\
  V_{n} \\
\end{bmatrix}
\label{ac_eq:LineAdmitMatrix_ji}
\end{equation}


A variety of symbols have been used to identify the two different line admittance matrices in \eqref{ac_eq:LineAdmitMatrix_ij} and \eqref{ac_eq:LineAdmitMatrix_ji}, e.g. $Y_{\text{line}}^{\text{from}}$, $Y_{\text{line}}^{\text{to}}$ in different textbooks, slides, or notes. In these lecture notes, we will use the symbols $\textbf{Y}_{\text{line},i \rightarrow j}$ to denote the line admittance matrix of \eqref{ac_eq:LineAdmitMatrix_ij} and $\textbf{Y}_{\text{line},j \rightarrow i}$ for the line admittance matrix of \eqref{ac_eq:LineAdmitMatrix_ji}.

\paragraph{Forming the Line Admittance Matrix}
\begin{enumerate}
 \item $\mathbf{\cmp{Y}_{\text{line}}}$ is an $L \times N$ matrix, where $L$ is the number of lines and $N$ is the number of nodes
 \item if row $k$ corresponds to line $i \rightarrow j$:
       \begin{itemize}
        \item start node $i$: $\cmp{Y}_{\text{line},ki}= \cmp{y}_{sh,i}+\cmp{y}_{ij}$
        \item end node $j$: $\cmp{Y}_{\text{line},kj}= -\cmp{y}_{ij}$
        \item rest of the row elements are zero
       \end{itemize}
 \item $\cmp{y}_{ij}= \dfrac{1}{R_{ij}+jX_{ij}}$ is the admittance of line $ij$
 \item $\cmp{y}_{sh,i}$ is the shunt capacitance $jB_{ij}/2$ of the $\pi$-model of the line
 \item We must create two $\mathbf{Y}_{\text{line}}$ matrices. One for $i \rightarrow j$ and one for $j\rightarrow i$.
\end{enumerate}

\subsection{Bus Admittance Matrix}
In order to be able to form the AC-OPF constraints, it is necessary to be able to compute the bus power injections. The bus power injection shows the net power that is entering or leaving a bus. By definition, in AC systems the net apparent power at a bus is equal to the product of the bus voltage (complex number) and the conjugate of the bus current (complex number):
\begin{equation}
 \cmp{S}_i = \cmp{V}_i \cmp{I}_i^*
\end{equation}

According to Kirchhoff's law, the net current injection at a bus is equal to the sum of currents leaving the bus. The sum of currents leaving a bus $i$ is the sum of the line currents flowing along all lines connected to bus $i$. This is expressed by \eqref{ac_eq:sum_currents}:


\begin{equation}
 \cmp{I}_i =\sum_k \cmp{I}_{ik}, \text{where $k$ are all the buses connected to bus $i$}
 \label{ac_eq:sum_currents}
\end{equation}

Let us now assume that from bus $i$ emanate two lines which connect $i$ to buses $m$ and $n$. Using also the result of \eqref{ac_eq:current_ij}, it is:
\begin{align}
 \cmp{I}_i & = \cmp{I}_{im}+ \cmp{I}_{in}   \nonumber   \\
           & = (\cmp{y}_{sh,i}^{i \rightarrow m}+\cmp{y}_{im})\cmp{V}_i - \cmp{y}_{im}\cmp{V}_m + (\cmp{y}_{sh,i}^{i \rightarrow n}+\cmp{y}_{in})\cmp{V}_i - \cmp{y}_{in}\cmp{V}_n \nonumber \\
           & = (\cmp{y}_{sh,i}^{i \rightarrow m}+\cmp{y}_{im}+\cmp{y}_{sh,i}^{i \rightarrow n}+\cmp{y}_{in})\cmp{V}_i - \cmp{y}_{im}\cmp{V}_m - \cmp{y}_{in}\cmp{V}_n
           \label{ac_eq:buscurrent}
\end{align}

In matrix form \eqref{ac_eq:buscurrent} is written as:
\begin{equation}
 \cmp{I}_i = [ y_{sh,im}+y_{im}+y_{sh,in}+y_{in} \quad -y_{im} \quad -y_{in}]
 \begin{bmatrix}
   V_i \\
   V_m\\
   V_n
 \end{bmatrix}
 \label{ac_eq:ybus_single}
\end{equation}


Similar to our considerations for the Line Admittance Matrix, and given our passion for vectors and matrices -- which must have become quite obvious by now (!) -- the question arises what is the algebraic relationship that can link vectors of bus currents and bus voltages. This is the role of the \emph{Bus Admittance Matrix} $\mathbf{Y}_{\text{bus}}$:

\begin{equation}
  \mathbf{I}_{\text{bus}} = \mathbf{Y}_{\text{bus}} \mathbf{V}
\end{equation}

Taking a close look at \eqref{ac_eq:ybus_single}, we observe that the matrix element multiplied with $V_i$ is the sum of all line admittances plus all the shunt elements connected to bus $i$ (usually these are the line shunts from the $\pi$-model, but it can also be additional shunt capacitors or inductors often used for reactive compensation). On the contrary, the matrix elements that are multiplied with the voltages at the neighboring buses are just the negative of the line admittance connecting the neighboring bus to bus $i$. For an arbitrary power system, where bus $i$ is connected to buses $m$ and $n$, bus 1 is only connected to bus 2, and bus $n$ is only connected to bus $i$, the Bus Admittance Matrix will have the following form:

\begin{equation}
\begin{bmatrix}
  I_{1} \\
  \vdots \\
  I_{i} \\
  \vdots \\
  I_{n} \\
\end{bmatrix}
=
\begin{bmatrix}
  \cmp{y}_{sh,1}+\cmp{y}_{12} & -\cmp{y}_{12} & 0 & \ldots & \ldots 0 \\
  \vdots & \vdots & \vdots & \vdots & \vdots  \\
  0\ldots & y_{sh,im}+y_{im}+y_{sh,in}+y_{in} & \ldots & -\cmp{y}_{im} & \ldots -\cmp{y}_{in} \\
  \vdots & \vdots & \vdots & \vdots & \vdots \\
  0\ldots & -\cmp{y}_{in} & \ldots & \ldots & \cmp{y}_{sh,n}+\cmp{y}_{in} \\
\end{bmatrix}
\begin{bmatrix}
  V_{1} \\
  V_2 \\
  \vdots \\
  V_{i} \\
  \vdots \\
  V_{n} \\
\end{bmatrix}
\label{ac_eq:BusAdmitMatrix}
\end{equation}

\paragraph{Forming the Bus Admittance Matrix}
\begin{enumerate}
 \item $\mathbf{\cmp{Y}_{\text{bus}}}$ is an $N \times N$ matrix, where $N$ is the number of nodes
 \item diagonal elements: $\cmp{Y}_{\text{bus},ii} = \sum_{t \in I} \cmp{y}_{sh,t}+\sum_k \cmp{y}_{ik}$, where $k$ are all the buses connected to bus $i$
 \item off-diagonal elements:
       \begin{itemize}
        \item $\cmp{Y}_{\text{bus},ij}= -\cmp{y}_{ij}$ if nodes $i$ and $j$ are connected by a line\footnote{If there are more than one lines connecting the same nodes, then they must all be added to $\cmp{Y}_{\text{bus},ij}, \cmp{Y}_{\text{bus},ii}, \cmp{Y}_{\text{bus},jj}$.}
        \item $\cmp{Y}_{\text{bus},ij}= 0$ if nodes $i$ and $j$ are not connected
       \end{itemize}
 \item $\cmp{y}_{ij}= \dfrac{1}{R_{ij}+jX_{ij}}$ is the admittance of line $ij$
 \item $\cmp{y}_{sh, i}$ are all shunt elements $t$ connected to bus $i$, including the shunt capacitance of the $\pi$-model of the line
\end{enumerate}

\subsection{AC Power Flow Equations}
From \eqref{ac_eq:BusAdmitMatrix}, it is $I_i =\cmp{\mathbf{Y}}_{\text{bus, row-i}} \cmp{\mathbf{V}}$, where $\cmp{\mathbf{Y}}_{\text{bus, row-i}}$ denotes the $i$-th row of the $\mathbf{Y}_{bus}$ matrix. Therefore, the apparent power at bus $i$ is:
\begin{align*}
 \cmp{S}_i & = \cmp{V}_i \cmp{I}_i^*                      \\
           & = \cmp{V}_i \cmp{\mathbf{Y}}_{\text{bus, row-i}}^* \cmp{\mathbf{V}}^*
\end{align*}

If we now want to express the bus apparent power in vector form, the problem is that we need to multiply from the left each row of the matrix $\mathbf{Y}_{\text{bus}}$ with the complex number $V_i$. To do that we introduce the notation $diag(\mathbf{V})$. Here, by $diag(\cmp{\mathbf{V}})$ we denote a diagonal $N \times N$ matrix, where the $N$ diagonal elements are equal to the $N \times 1$ vector $\cmp{\mathbf{V}}$, and all the rest of the matrix elements are zero.

Then, for all buses, the apparent power $\cmp{\mathbf{S}}= [\cmp{S}_1 \ldots \cmp{S}_N]^T$ is:
\begin{equation}
 \mathbf{S} = diag(\cmp{\mathbf{V}}) \cmp{\mathbf{Y}}_{\text{bus}}^* \cmp{\mathbf{V}}^*
 \label{ac_eq:acpowerflow}
\end{equation}

The net apparent power injection at every bus is equal to the total generation minus the total load connected at the bus. In vector form:
\begin{equation}
  \mathbf{S} = \mathbf{S}_{\text{gen}} - \mathbf{S}_{\text{load}}
\label{ac_eq:SgenSload}
\end{equation}
Combining \eqref{ac_eq:acpowerflow} and \eqref{ac_eq:SgenSload}, the AC power flow equations are given by:
\begin{equation}
 \mathbf{S}_{\text{gen}} - \mathbf{S}_{\text{load}} = diag(\cmp{\mathbf{V}}) \cmp{\mathbf{Y}}_{\text{bus}}^* \cmp{\mathbf{V}}^*
\end{equation}

\section{Formulation of the optimization problem}

In Section~\ref{ac_sec:acpowerflow}, we have introduced two types of equations. First, the equations about the line currents. Second, the equations about the bus current and bus power injections. These two sets of equations will form a fundamental part of our constraints in the AC Optimal Power Flow problem:

\begin{equation}
   \min_{P_{G_i}} \pmb{c}^T \pmb{P}_{G}
   \label{ac_eq:objfun}
\end{equation}
subject to:
 \begin{align}
  \text{\color{blue}{AC flow}}           \quad                      & \mathbf{S}_G-\mathbf{S}_L=diag(\cmpx{\mathbf{V}})\cmpx{\mathbf{Y}}_{\text{bus}}^*\cmpx{\mathbf{V}}^* \label{ac_eq:nodalequations}\\
  \text{\color{blue}{Line Current}}        \quad                    & |\cmpx{\mathbf{Y}}_{\text{line}, \textcolor{red}{i \rightarrow j}}\cmpx{\mathbf{V}}| \leq \mathbf{I}_{line,max} \label{ac_eq:currentineq_ij}\\
                                                                    & |\cmpx{\mathbf{Y}}_{\text{line}, \textcolor{red}{j \rightarrow i} }\cmpx{\mathbf{V}}| \leq \mathbf{I}_{line,max} \label{ac_eq:currentineq_ji} \\
  \text{\color{red}{\it or }\color{blue}{Apparent Flow}}      \quad & | \cmpx{V}_{\textcolor{red}{i}} \cmpx{\mathbf{Y}}_{\text{line}, \textcolor{red}{i \rightarrow j},\textcolor{red}{\text{row-i}}}^{*}\cmpx{\mathbf{V}}^*| \leq S_{\textcolor{red}{i \rightarrow j},max} \label{ac_eq:powerineq_ij} \\
                                                                    & | \cmpx{V}_{\textcolor{red}{j}} \cmpx{\mathbf{Y}}_{\text{line}, \textcolor{red}{j \rightarrow i},\textcolor{red}{\text{row-j}}}^{*}\cmpx{\mathbf{V}}^*| \leq S_{\textcolor{red}{j \rightarrow i}, max} \label{ac_eq:powerineq_ji} \\
  \text{\color{blue}{Gen. Active Power}}  \quad                     & \mathbf{0} \leq \mathbf{P}_G \leq \mathbf{P}_{G,max} \label{ac_eq:activepowerbounds}\\
  \text{\color{blue}{Gen. Reactive Power}}\quad                     & -\mathbf{Q}_{G,max} \leq \mathbf{Q}_G \leq \mathbf{Q}_{G,max} \label{ac_eq:reactivepowerbounds}\\
  \text{\color{blue}{Voltage Magnituge}}  \quad                     & \mathbf{V}_{min} \leq \mathbf{V} \leq \mathbf{V}_{max} \label{ac_eq:voltagebounds} \\
  \text{\color{blue}{Voltage Angle}}       \quad                    & \pmb{\delta}_{min} \leq \pmb{\delta} \leq \pmb{\delta}_{max} \label{ac_eq:deltabounds}
 \end{align}

All shown variables are vectors or matrices. The bar above a variable denotes complex numbers. The operator $(\cdot)^*$ denotes the complex conjugate. To simplify notation, the bar denoting a complex number is dropped in the rest of this chapter. Attention! The {\it current} flow constraints are defined as {\it vectors}, i.e. for all lines. The apparent power {\it line} constraints are defined {\it per line}.

The AC-OPF formulation, as we present it in \eqref{ac_eq:objfun}, minimizes the active power generation costs.
However, for the AC-OPF, we can set several different objectives. For a more detailed discussion, please refer to Section~\ref{ac_sec:acopfobjfun}.

Constraints \eqref{ac_eq:nodalequations} are the only equality constraints in the standard AC-OPF formulation, and they denote the AC power flow equations. Any operating point that will be determined through the optimization must satisfy the AC power flow equations in order to be a true operating point of the power system.

Constraints \eqref{ac_eq:currentineq_ij} - \eqref{ac_eq:currentineq_ji} represent the line current inequality constraints. If our transmission line limits are represented by the current thermal limit, these are the inequality constraints that must be used. If, on the other hand, the line thermal limits are set by the apparent power, then \eqref{ac_eq:powerineq_ij} -- \eqref{ac_eq:powerineq_ji} must be used, which refer to the apparent power flow.
\emph{In any case, only one set of these constraints must be used:} either the line current limit \eqref{ac_eq:currentineq_ij} - \eqref{ac_eq:currentineq_ji} of the apparent power line flow limit \eqref{ac_eq:powerineq_ij} -- \eqref{ac_eq:powerineq_ji}.

Constraints \eqref{ac_eq:activepowerbounds} refer to the active power bounds of the generators, while constraints \eqref{ac_eq:reactivepowerbounds} refer to the generators' reactive power bounds. Constraints \eqref{ac_eq:voltagebounds} refer the maximum and minimum allowable limits for the bus voltage magnitudes (no complex numbers here), and constraints \eqref{ac_eq:deltabounds} refer to maximum and minimum allowable limits for the voltage angles.

\subsection{Objective Function for the AC Optimal Power Flow}
\label{ac_sec:acopfobjfun}
Despite the goal to clear all electricity markets with an algorithm based on the AC-OPF in the future, the AC Optimal Power Flow can be used for a number of different purposes. Here we list four of those, among several examples.

\paragraph{Minimization of costs:} As shown in \eqref{ac_eq:objfun}, one of the most common uses for the AC-OPF is the minimization of the costs for producing electricity. The objective function minimizes the total costs for generating active power:
\begin{equation}
   \min_{P_{G_i}} \pmb{c}^T \pmb{P}_{G}
   \label{ac_eq:objfun}
\end{equation}
This objective function can be used both for clearing electricity markets, but also in vertically integrated utilities that want to reduce their operating costs (in that case possibly assuming quadratic costs).

\paragraph{Minimization of active and reactive power losses:} The minimization of the active and reactive power losses is probably among the most important daily functions of the system operator -- after, of course, ensuring that the power system is secure and not threatened by blackouts. To minimize the active and reactive power losses, the objective function can be of the form:
\begin{equation}
  \min \sum_{i,j \in L} S_{ij}+S_{ji}
\end{equation}

\begin{figure}[!h]
  \centering
   \begin{tikzpicture}
    \draw (0,-0.3) -- (0,0.3);
    \draw (0,0) -- (6,0);
    \draw (6,-0.3) -- (6,0.3);
    \draw (1,0) -- (1,-2.5);
    \draw (5,0) -- (5,-2.5);
    \draw (0.8,-2.5) -- (1.2,-2.5);
    \draw (0.85,-2.55) -- (1.15,-2.55);
    \draw (0.9,-2.6) -- (1.1,-2.6);
    \draw (5,0) -- (5,-2.5);
    \draw (4.8,-2.5) -- (5.2,-2.5);
    \draw (4.85,-2.55) -- (5.15,-2.55);
    \draw (4.9,-2.6) -- (5.1,-2.6);
    \node [above] at (0,0.3) {$i$};
    \node [above] at (6,0.3) {$j$};
    \node [below] at (3, -0.1) {$R_{ij}+jX_{ij}$};
    \node [right] at (1, -1.5) {$\dfrac{jB_{ij}}{2}$};
    \node [left] at (5, -1.5) {$\dfrac{jB_{ij}}{2}$};
    \draw [thick, ->] (0,0) -- (0.7,0);
    \draw [thick, ->] (5.3,0) -- (6,0);
    \node [above] at (0.7,0) {$S_{i \rightarrow j}$};
    \node [above] at (5.3,0) {$-S_{j \rightarrow i}$};
   \end{tikzpicture}
  \caption{$\pi$-model of the line and apparent power flows}
  \label{ac_fig:pimodellosses}
 \end{figure}
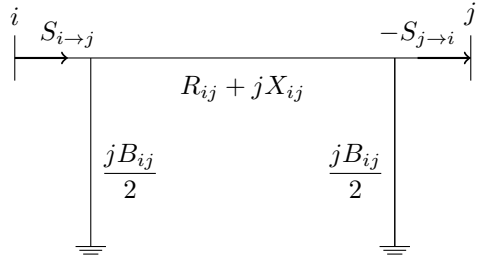

The total losses along a line $i - j$ are equal to the apparent power leaving node $i$ minus the apparent power arriving node $j$. As shown in Fig.~\ref{ac_fig:pimodellosses}, the power leaving node $i$ is $S_{i \rightarrow j}$. The apparent power arriving at node $j$ is equal to $-S_{j \rightarrow i}$; this is because by convention the direction $j \rightarrow i$ is positive, so the power exiting node $j$ will have a direction opposite to $j \rightarrow i$.

According to this:
\begin{align}
 S_{\text{losses}} & = S_{\textcolor{red}{i \rightarrow j}} - (- S_{\textcolor{red}{j \rightarrow i}}) \\
 S_{\text{losses}} & = S_{\textcolor{red}{i \rightarrow j}} \textcolor{red}{+} S_{\textcolor{red}{j \rightarrow i}}
\end{align}

\paragraph{Maintaining a constant voltage profile:} At times power system operators may want to maintain a constant voltage profile across all or a part of the system nodes. This, for example, might help them avoid voltage instability problems. An optimization in this case may help them identify what is the preferrable set of actions, usually related to the injection or absorption of reactive power, to achieve the desired profile. The objective function can have for example the following form (among several possibilities):
\begin{equation}
\min \sum_i (V_i-V_{\text{setpoint},i})^2
\end{equation}
The quadratic objective function in this case helps us minimize both the positive and the negative deviations of the voltage from the desired setpoint. A different option would be to minimize the absolute value of the deviation, i.e. $\min \sum_i |V_i-V_{\text{setpoint},i}|$. In that case, however, for most solvers we need to reformulate our problem in order to remove the absolute value from our objective function, as it is a non-smooth function.

For this problem, we may also consider different objective functions, e.g. to minimize both the voltage deviation and the required changes in reactive power -- but we must be careful when it comes to multi-objective optimization. We can also consider additional constraints for our problem, e.g. that the active power injection at every bus must remain constant to the pre-decided level. In that case we have to replace \eqref{ac_eq:activepowerbounds} with an equality constraint of the form $\pmb{P}=\pmb{P}_\text{setpoint}$.

\paragraph{Transmission investments:} This formulation usually involves mixed integer programming, as the optimizer must be able to choose among different options for new transmission lines. In its simplest form, however, the goal of transmission investments might be to determine what is the minimum capacity upgrades in existing lines, in order to reduce the operating cost. To simplify things further, as an initial ``back-of-the-envelope'' calculation one may neglect that the line upgrades are performed in blocks, i.e. either upgrading an exisiting line by e.g. 100 MVA or not; upgrading to any desirable value i.e. 53.5 MVA is impossible -- not from a technical perspective, but rather from the availability of the options provided by manufacturers. To avoid this discontinuity, one can assume that the line upgrade is a continuous variable. In that case, the objective function will look like:
\begin{equation}
\min \sum_{t \in H} \mathbf{c}_t^T \mathbf{P}_G,t + \mathbf{c}_{\text{line}}^T \mathbf{s}_{\text{line}}
\end{equation}
where $\sum_{t \in H} \mathbf{c}_t^T \mathbf{P}_G,t$ is the sum of the generating costs over the projected lifetime of the transmission line, $\mathbf{c}_{\text{line}}$ is the cost of line upgrade per unit of additional capacity, and $\mathbf{s}_{\text{line}}$ is the newly added capacity, measured either in Amperes (current) or in MVA (power).

Besides the objective function, we also have to add variable $\mathbf{s}_{\text{line}}$ either to line current inequalities \eqref{ac_eq:currentineq_ij} - \eqref{ac_eq:currentineq_ji} of the apparent power flow \eqref{ac_eq:powerineq_ij} -- \eqref{ac_eq:powerineq_ji}. For example, for the line current inequalities, this would look like:
\begin{equation}
|\cmpx{\mathbf{Y}}_{\text{line}}\cmpx{\mathbf{V}}| \leq \mathbf{I}_{line,max} + \mathbf{s}_{\text{line}}
\end{equation}
After we solve the optimization, we can then perform a more detailed cost-benefit analysis and select one of the available capacity upgrade options, that will probably be close to our optimization result.

\subsection{Transmission Line Limit constraints: Current vs Power}
As shown in  \eqref{ac_eq:currentineq_ij} - \eqref{ac_eq:currentineq_ji} and \eqref{ac_eq:powerineq_ij} -- \eqref{ac_eq:powerineq_ji}, we can have different types of inequalities to express the line limits. But what are the inequalities we should use, and when each of them is most appropriate?

Several OPF formulations are imposing the apparent power limits for overhead lines, cables, and transformers. This should not be always the case, however.

Especially for overhead lines (OHL) and cables, it is preferrable to use the \emph{line current limits}. The OHL and cable manufacturers usually indicate the thermal limits of their conductors in Amperes, since it is the current that is the limiting factor for the thermal stress of the OHL or cables. So, for overhead lines and cables it is better to use the line current inequalities \eqref{ac_eq:currentineq_ij} - \eqref{ac_eq:currentineq_ji} as limits.

For transformers, on the other hand, it is difficult to set the current limit as constraint, as primary side and secondary side have different line currents (although in the per unit system they will both be the same). The manufacturers of transformers usually give the apparent power as the limit to prevent overloads and excessive thermal stress of their equipment. So, for transformers, it is usually preferrable to use the apparent power limit, as expressed by \eqref{ac_eq:powerineq_ij} -- \eqref{ac_eq:powerineq_ji}.

Besides current and apparent power limits, literature also suggests the use of active power limits. Indeed, active power can be the limiting factor for the transmission along very long lines. Beyond certain lengths, and as the reactance $x_{ij}$ of the line increases, the limiting factor for the power transfer becomes its steady-state stability limit, which is given by the relationsip:
\begin{equation}
  P_{ij, max} = \frac{1}{x_{ij}}V_i V_j
\end{equation}
In case of very long lines (over a few hundred of kilometers) using the active power flow limit is preferrable. In that case, the active power flow along a line will be equal to $\Re(S_{i \rightarrow j}) = \Re(V_i I_{i \rightarrow j}^*)$. Following the derivations we present in Section~\ref{ac_sec:opcomplexnum}, the active power limit inequality constraints are given by the equations:
\begin{equation}
  \begin{bmatrix}
    \Re(V_i) & -\Im(V_i)
  \end{bmatrix}
  \begin{bmatrix}
    \Re(\mathbf{Y}_{\text{line}, i \rightarrow j}) &-\Im(\mathbf{Y}_{\text{line}, i \rightarrow j}) \\
    \Im(\mathbf{Y}_{\text{line}, i \rightarrow j}) &\quad \Re(\mathbf{Y}_{\text{line}, i \rightarrow j})
  \end{bmatrix}^*
  \begin{bmatrix}
  \Re(\mathbf{V}) \\
  \Im(\mathbf{V})
  \end{bmatrix}^*
  \leq P_{ij,max}
\end{equation}

\begin{equation}
  \begin{bmatrix}
    \Re(V_j) & -\Im(V_j)
  \end{bmatrix}
  \begin{bmatrix}
    \Re(\mathbf{Y}_{\text{line}, j \rightarrow i}) &-\Im(\mathbf{Y}_{\text{line}, j \rightarrow i}) \\
    \Im(\mathbf{Y}_{\text{line}, j \rightarrow i}) &\quad \Re(\mathbf{Y}_{\text{line}, j \rightarrow i})
  \end{bmatrix}^*
  \begin{bmatrix}
  \Re(\mathbf{V}) \\
  \Im(\mathbf{V})
  \end{bmatrix}^*
  \leq P_{ji,max}
\end{equation}

Notice that $\Re(V_i), -\Im(V_i), \Re(V_j), -\Im(V_j), P_{ij,max}$ and $P_{ji,max}$ are scalars while the rest, shown in boldface, are vectors or matrices.

\paragraph{Line Limits:} Absent of real models and datasheets, we often need to use publicly available power system models. These sources may often contain only the apparent power flow limit and provide no information about the line current limit (e.g. this is the case for Matpower case files, and some IEEE systems). There is a straightforward way to transform any apparent power limit to a good approximation of an equivalent line current limit. Here is how to do this.

As it must be well known by now, it is: $S=VI^*$. Taking the absolute value of the left and right terms, and considering that for any complex number holds $|z|=|z^*|$, it is:
\begin{equation}
|S|=|V||I|
\label{ac_eq:svi_limit}
\end{equation}
Assuming that our lines are operated at nominal voltage (this is the only approximation we make), it must be $V=1 \text{p.u.}$ So, as long as \eqref{ac_eq:svi_limit} is expressed in \emph{per unit}, then $|S|=|I|$. This means that the line current limit is equal to the apparent power limit expressed in per unit:
\begin{equation}
S_{i \rightarrow j, max} = I_{i \rightarrow j, max} \; \text{in per unit;  or} \quad  \frac{S_{i \rightarrow j, max}}{S_{\text{base}}}=\frac{I_{i \rightarrow j, max}}{I_{\text{base}}}
\end{equation}

\subsection{Bus Voltage Magnitude Limits}
To maintain safe operation, we usually bound the voltage magnitude within a small range of aceeptable limits. In the vast majority of cases we allow a maximum voltage deviation of 10\% or less. This means that the voltage bounds are often \mbox{$V_{\text{min}}=0.9 \, \text{p.u.}$} and $V_{\text{max}}=1.1 \, \text{p.u.}$ However, we can often impose even tighter limits, e.g. \mbox{$V_{\text{min}}=0.95 \, \text{p.u.}$} and $V_{\text{max}}=1.05 \, \text{p.u.}$ In general, any voltage magnitude limits that do not allow more than 10\% deviation are acceptable, but depending on the system and the application, we need to be considerate of tighter voltage limits.

\subsection{Bus Voltage Angle Limits}
Voltage angle limits are important for the non-linear solver of the AC-OPF. Due to the fact that our equations use complex numbers, we can obtain several instances of exactly the same power flow solution in ``angle intervals'' of $360^{o}$. For example, $V=1.02 \angle{15^{o}} \, = \, 1.02 \angle{375^{o}} \, = \, 1.02 \angle{-345^{o}}$.
By bounding our bus voltage angles within the range of $0^{o}$ and $360^{o}$, we guide the solver to a unique solution and avoid possible degenerate cases, where the solver oscillates between two exactly equivalent solutions that are just shifted by $360^{o}$. Limiting our solutions between $-180^{o}$ and $180^{o}$ is equivalent to 0 and $360^{o}$ and is usually preferrable. Since, in our formulations we are usually computing in rad, then the most usual angle limits are:

\begin{equation}
\delta_{min} = -\pi \leq \delta \leq \pi = \delta_{max}
\end{equation}

\section[Operations with complex numbers]{Operations with complex numbers when formulating the optimization problem}
\label{ac_sec:opcomplexnum}
Non-linear solvers in Matlab accept complex numbers, so we can directly enter the problem formulation in a complex number format. However, there are several other solvers that cannot deal with complex numbers, and require the input of real numbers instead. Interfaces like YALMIP overcome this issue for the user by directly accepting the complex number formulation and translating it to real numbers\todo{ask Andreas if this is true}.

Being able to transform complex number operations to operations with real numbers is a skill that is sometimes required (e.g. if we code in a different environment from Matlab, or dealing with semidefinite programming, as we will see in the next chapter). In this section we describe how we can deal with this problem.

Suppose we want to multiply complex numbers $z_1 = a+jb$ and $z_2 = c+jd$. Then:
\begin{equation}
  z_3 = z_1 z_2 = (a+jb)(c+jd) = ac-bd+j(ad+bc)
\end{equation}
We can separate the real and imaginary part of a complex number by treating $z$ as a $2 \times 1$ vector of real numbers, in the form: $z= [ a\quad b]^T$.

Having this in mind, we can transform the multiplication of two complex numbers to a multiplication of a matrix with a vector of real numbers, as follows:
\begin{equation}
  z_3 =
\begin{bmatrix}
  ac-bd \\
  bc+ad
\end{bmatrix}
= \begin{bmatrix}
  a & -b \\
  b & \; a
\end{bmatrix}
\begin{bmatrix}
c \\
d
\end{bmatrix}
\end{equation}

In other words, we can express the multiplication of any complex numbers to a multiplication of real numbers as follows:
\begin{equation}
  \begin{bmatrix}
  \Re(z_3) \\
  \Im(z_3)
  \end{bmatrix}
  =
  \begin{bmatrix}
    \Re(z_1) & -\Im(z_1) \\
    \Im(z_1) & \quad \Re(z_1)
  \end{bmatrix}
  \begin{bmatrix}
  \Re(z_2) \\
  \Im(z_2)
  \end{bmatrix}
\end{equation}

This property also extends to multiplication of vectors of complex numbers. Assume $\mathbf{I}=\mathbf{YV}$, where $\mathbf{I}, \mathbf{V}$ are complex vectors, and $\mathbf{Y}$ is a complex matrix. From the above equation, we can generalize as follows:
\begin{equation}
  \begin{bmatrix}
    \Re(\mathbf{I}) \\
    \Im(\mathbf{I})
  \end{bmatrix}
=
\begin{bmatrix}
  \Re(\mathbf{Y}) &-\Im(\mathbf{Y}) \\
  \Im(\mathbf{Y}) &\quad \Re(\mathbf{Y})
\end{bmatrix}
\begin{bmatrix}
\Re(\mathbf{V}) \\
\Im(\mathbf{V})
\end{bmatrix}
\end{equation}

\section{From the AC power flow equations to the DC power flow}
The time for your patience since Chapter~\ref{chap:dcopf} and \eqref{dc_eq:activepowerAC} is now coming to an end! Starting from the full AC power flow equations, in this section we will present how we arrived at equation \eqref{dc_eq:activepowerAC}, which formed the basis of our DC power flow equations. This creates the link between AC power flow and DC power flow, while explicitly stating all assumptions made.

The power flow along a line is:

\begin{equation}
 S_{i \rightarrow j} = V_i I_{i \rightarrow j}^* = V_i (y_{sh,i}^*V_i^* +y_{ij}^*(V_i^* - V_j^*))
 \label{ac_eq:acdcpowerflow1}
\end{equation}

Assumptions:
\begin{enumerate}
 \item Assume a negligible shunt conductance: $g_{sh,ij}=0 \Rightarrow y_{sh,i}=jb_{sh,i}$. This is true for most power system modeling approaches, as even the $\pi$-model of a transmission line usually neglects the shunt conductance.
 \item Assume a negligible series resistance: $z_{ij}=r_{ij}+jx_{ij}\approx jx_{ij}$. Then $y_{ij}=-j\dfrac{1}{x_{ij}}$. In \emph{transmission} systems it is usually $R<<X$ for the series impedance in the $\pi$-model of the transmission line. Therefore, especially for lightly loaded systems, neglecting it will not have a large impact on the results. (Note: this assumption is not quite valid in distribution systems, where $R$ can be comparable to $X$; the full AC power flow equations must be used for optimization in distribution systems)
\end{enumerate}

According to the above two assumptions, the current $I_{i \rightarrow j}^*$, shown in \eqref{ac_eq:acdcpowerflow1}, becomes:
\begin{equation}
	I_{ij}^* = -jb_{sh,i}V_i^* +j\frac{1}{x_{ij}}(V_i^* - V_j^*))
\end{equation}

\begin{enumerate}
 \item[3.] Assume: $V_i = V_i \angle{0}$ and $V_j = V_j \angle{\theta}$, with $\theta = \delta_j - \delta_i$ (no loss of generality with this assumption). Then:
\end{enumerate}
\begin{align}
 I_{ij}^*  = & -jb_{sh,i}V_i+j\frac{1}{x_{ij}}(V_i - (V_j \cos \theta -j V_j \sin \theta))                           \\
 =           & -jb_{sh,i}V_i+j\frac{1}{x_{ij}}V_i - j \frac{1}{x_{ij}}V_j \cos \theta -\frac{1}{x_{ij}}V_j\sin\theta
\end{align}

As $V_i = V_i \angle{0}$, $V_i$ is a real number. Then the active power transfer along a line is:

\begin{align}
 P_{ij} = & \Re\{S_{ij}\} = V_i \Re\{I_{ij}^*\} = -\frac{1}{x_{ij}} V_i V_j \sin \theta
\end{align}

As we have set $\delta = \theta_j - \theta_i$, it is:

\begin{align}
 P_{ij} = \frac{1}{x_{ij}} V_i V_j \sin (\delta_i - \delta_j),
\end{align}
which is exactly the equation \eqref{dc_eq:activepowerAC}

As already mentioned in Chapter~\ref{chap:dcopf}, to arrive at the DC power flow equations, we further make the following assumptions:
       \begin{enumerate}
        \item[4.] $V_i$, $V_j$ are constant and equal to 1 p.u.
        \item[5.] $\sin \theta \approx \theta$, where $\theta$ must be in rad
      \end{enumerate}

Then:
\begin{center}
 $P_{ij} = \dfrac{1}{x_{ij}} (\delta_i - \delta_j)$
\end{center}

For a more detailed discussion of the assumptions (4) and (5), the interested reader is referred to Chapter~\ref{chap:dcopf}.

\section{Nodal Prices (LMPs) in AC-OPF}
The nodal prices in the AC-OPF are the lagrangian multipliers of the equality constraints for the \emph{active power flow}.

To calculate them, we need to split equality constraints \eqref{ac_eq:nodalequations} to their real and imaginary part, as follows:
\begin{align}
  \mathbf{P}_G-\mathbf{P}_L &= \Re{diag(\cmpx{\mathbf{V}})\cmpx{\mathbf{Y}}_{\text{bus}}^*\cmpx{\mathbf{V}}^*} \label{ac_eq:nodalequations_P}\\
  \mathbf{Q}_G-\mathbf{Q}_L &= \Im{diag(\cmpx{\mathbf{V}})\cmpx{\mathbf{Y}}_{\text{bus}}^*\cmpx{\mathbf{V}}^*} \label{ac_eq:nodalequations_Q}
\end{align}

The Lagrangian multipliers associated with equality constraints \eqref{ac_eq:nodalequations_P} are the nodal prices for the AC-OPF problem. They show what is the marginal cost of the demand of an additional MW at each node.
\subsection{LMPS in the AC-OPF vs DC-OPF}
\textbf{Attention:} The nodal prices calculated through the AC-OPF will \emph{always} differ at different buses. Contrary to the DC-OPF, where the nodal prices are exactly the same at all buses if there is no congestion, this is not the case here. The AC-OPF explicitly considers the line losses, and these are an integral component of the nodal prices in the AC-OPF. As a result, besides the generator cost and the contribution to the line congestion, in the AC-OPF there is a third cost component which has to do with the amount of losses incurred to the system by the demand of an extra MW at a specific bus.

Considering that the consumption of an additional MW will incur a different amount of system losses depending on where this is consumed, this is why the LMPs in the AC-OPF always differ from each other, even if there is no line congestion.


\appendix

\chapter{Derivation of Locational Marginal Prices}
\label{chap:appendix_LMP}
\begin{equation}
 \min \sum_i c_i P_{G_i} \notag
\end{equation}
subject to:
\begin{equation}
 P_{G_i}^{min} \leq   P_{G_i} \leq P_{G_i}^{max} \notag
\end{equation}
\begin{equation}
 \sum_i P_{G_i} -  \sum_i P_{D_i} = 0 \notag
\end{equation}
\begin{equation}
 {\bf PTDF} ({\bf P_G} - {\bf P_D}) \leq {\bf P_{line}^{max}} \notag
\end{equation}

\begin{align}
 \mathcal{L}(P_G, \nu, \lambda, \mu) =\sum_{i=1}^{N_{P_G}} c_i P_{G,i} + \nu \cdot \left( \sum_{i=1}^{N_{P_G}} P_{G,i} - \sum_{i=1}^{N_{P_L}} P_{L,i} \right) \nonumber \\
 + \sum_{i=1}^{N_{L}} \lambda_i^{+} \cdot \left[ \mathbf{PTDF_i} \cdot (\mathbf{P_G} - \mathbf{P_L})-F_{L,i} \right] \nonumber                                          \\
 + \sum_{i=1}^{N_{L}} \lambda_i^{-} \cdot \left[-\mathbf{PTDF_i} \cdot (\mathbf{P_G} - \mathbf{P_L})-F_{L,i} \right] \nonumber                                          \\
 + \sum_{i=1}^{N_{P_G}} \mu_i^{+} \cdot (P_{G,i}-P_{G,i,max}) + \sum_{i=1}^{N_{P_G}} \mu_i^{-} \cdot (-P_{G,i}) \notag
 \label{eq_51}
\end{align}

\begin{itemize}
 \item Assume a 3-bus system with 3 generators, and 1 load on bus 3
 \item We assume an auxilliary variable $\xi_3$ that represents very small changes of the load in Bus 3. We assume $\xi_3=0$.
 \item Then it is $\hat{P_L} = P_L + \Xi$, where $\Xi=[0 \; 0 \; \xi_3]^T$.
\end{itemize}
\begin{figure}[!h]
 \centering
 \resizebox{6cm}{!} {
  \begin{tikzpicture}
   \node [above] at (1,0.5) {$1$};
   \node [above] at (5,0.5) {$2$};
   \node [left] at (2.3, -3) {$3$};
   \draw (0,0) circle (0.5cm);
   \draw (0.5,0) -- (1,0);
   \draw (1,-1) -- (1,0.5);
   \draw (1,0) -- (5,0);
   \draw (5,-1) -- (5,0.5);
   \draw (5, 0) -- (5.5,0);
   \draw (6,0) circle (0.5cm);
   \draw (1,-0.5) -- (1.5,-0.5);
   \draw (1.5,-0.5) -- (2.7,-2.5);
   \draw (2.7,-2.5) -- (2.7,-3);
   \draw (5,-0.5) -- (4.5,-0.5);
   \draw (4.5,-0.5) -- (3.3,-2.5);
   \draw (3.3,-2.5) -- (3.3,-3);
   \draw (2.3,-3) -- (3.7,-3);
   \draw [line width=0.4mm,->] (2.5,-3) -- (2.5,-4);
   \draw (3.5,-3) -- (3.5,-3.5);
   \draw (3.5,-4) circle (0.5cm);
   \draw (-0.3,0) sin (-0.15,0.2) cos (0,0) sin (0.15,-0.2) cos (0.3,0);
   \draw (5.7,0) sin (5.85,0.2) cos (6,0) sin (6.15,-0.2) cos (6.3,0);
   \draw (3.2,-4) sin (3.35,-3.8) cos (3.5,-4) sin (3.65,-4.2) cos (3.8,-4);
  \end{tikzpicture}
 }
\end{figure}

\begin{align}
 \mathcal{L}(P_G, \nu, \lambda, \mu, \Xi) =\sum_{i=1}^{N_{P_G}} c_i P_{G,i} + \nu \cdot \left( \sum_{i=1}^{N_{P_G}} P_{G,i} - \sum_{i=1}^{N_{P_L}} P_{L,i} -\xi_i\right) \nonumber \\
 + \sum_{i=1}^{N_{L}} \lambda_i^{+} \cdot \left[ \mathbf{PTDF_i} \cdot (\mathbf{P_G} - \mathbf{P_L}-\Xi)-F_{L,i} \right] \nonumber                                                         \\
 + \sum_{i=1}^{N_{L}} \lambda_i^{-} \cdot \left[-\mathbf{PTDF_i} \cdot (\mathbf{P_G} - \mathbf{P_L}-\Xi)-F_{L,i} \right] \nonumber                                                         \\
 + \sum_{i=1}^{N_{P_G}} \mu_i^{+} \cdot (P_{G,i}-P_{G,i,max}) + \sum_{i=1}^{N_{P_G}} \mu_i^{-} \cdot (-P_{G,i}). \notag
\end{align}

\begin{itemize}
 \item To save space in this slide: $K_{i} \equiv PTDF_{i}$
\end{itemize}
\begin{align}
 \mathcal{L}(P_G, \nu, \lambda, \mu, \xi_3) =\sum_{i=1}^{N_{P_G}} c_i P_{G,i} + \nu \cdot \left( \sum_{i=1}^{N_{P_G}} P_{G,i} - \sum_{i=1}^{N_{P_L}} P_{L,i} -\xi_3\right) \nonumber \\
 + \sum_{i=1}^{N_{L}} \lambda_i^{+} \cdot \left[ K_{i,1} \cdot P_{G,1} + K_{i,2} \cdot P_{G,2} + K_{i,3} \cdot (P_{G,3} -P_{L,3} - \xi_3)-F_{L,i} \right] \nonumber                          \\
 + \sum_{i=1}^{N_{L}} \lambda_i^{-} \cdot \left[ -K_{i,1} \cdot P_{G,1} - K_{i,2} \cdot P_{G,2} - K_{i,3} \cdot (P_{G,3} -P_{L,3} - \xi_3)-F_{L,i} \right] \nonumber                         \\
 + \sum_{i=1}^{N_{P_G}} \mu_i^{+} \cdot (P_{G,i}-P_{G,i,max}) + \sum_{i=1}^{N_{P_G}} \mu_i^{-} \cdot (-P_{G,i}). \notag
\end{align}

\begin{itemize}
 \item No congestion $\Rightarrow$ all $\lambda_i=0$
 \item One marginal generator: only one generator has both $\mu_i^+=0$ and $\mu_i^-=0$
 \item Assume G2 is marginal; $P_{G1}=P_{G1,max}$; $P_{G3}=0$.
\end{itemize}
\begin{minipage}{0.48\textwidth}
 \begin{align}
  \frac{\partial \mathcal{L}}{\partial P_{G,i}} & =0, \quad \text{for all } i\in N_{P_G} \notag \\
                                                & c_1+\nu+\mu_1^+ =0 \notag                     \\
                                                & c_2+\nu  =0       \notag                      \\
                                                & c_3+\nu+\mu_3^- =0 \notag
 \end{align}
\end{minipage}
\begin{minipage}{0.48\textwidth}
 Marginal change in the cost function for a marginal change in load:
 \begin{equation}
  LMP_3 = \frac{\partial \mathcal{L}}{\partial \xi_3} = -\nu \notag
 \end{equation}
\end{minipage}
\begin{center}
 Attention! $\xi_3$ does not exist in the optimization problem and is not an optimization variable. We do not need to derive any KKT conditions w.r.t. $\xi_3$, e.g. $\frac{\partial L}{\partial \xi_3} = 0$. \\ $\xi_3$ is just an auxilliary variable. It helps us ``represent'' the marginal change in the load of bus 3. $\frac{\partial L}{\partial \xi_3}$ quantifies its effect on the Lagrangian.
\end{center}

\begin{itemize}
 \item No congestion $\Rightarrow$ all $\lambda_i=0$
 \item One marginal generator: only one generator has both $\mu_i^+=0$ and $\mu_i^-=0$
 \item Assume G2 is marginal; $P_{G1}=P_{G1,max}$; $P_{G3}=0$.
\end{itemize}
\begin{minipage}{0.48\textwidth}
 \begin{align}
  \frac{\partial \mathcal{L}}{\partial P_{G,i}} & =0, \quad \text{for all } i\in N_{P_G} \notag \\
                                                & c_1+\nu+\mu_1^+ =0 \notag                     \\
                                                & c_2+\nu  =0       \notag                      \\
                                                & c_3+\nu+\mu_3^- =0 \notag
 \end{align}
\end{minipage}
\begin{minipage}{0.48\textwidth}
 Marginal change in the cost function for a marginal change in load:
 \begin{equation}
  LMP_3 = \frac{\partial \mathcal{L}}{\partial \xi_3} = -\nu \notag
 \end{equation}
\end{minipage}
\begin{center}
 $LMP_3 = -\nu = c_2$: nodal price on bus 3!\\
 How much is the LMP on the other buses?
\end{center}

\begin{itemize}
 \item Assume that line 1-3 gets congested in the direction $1 \rightarrow 3 \Rightarrow \lambda_{13}^+ \neq 0$
 \item Now G2 and G3 are both marginal gens; $P_{G1}=P_{G1,max}$.
\end{itemize}
\begin{minipage}{0.48\textwidth}
 \begin{align}
    & \frac{\partial \mathcal{L}}{\partial P_{G,i}}  =0, \quad \text{for all } i\in N_{P_G} \notag \\
    & c_1+\nu+\mu_1^+ +\lambda_{13}^+ PTDF_{13,1}=0 \notag                                         \\
    & c_2+\nu +\lambda_{13}^+ PTDF_{13,2} =0       \notag                                          \\
    & c_3+\nu+\lambda_{13}^+ PTDF_{13,3} =0 \notag
 \end{align}
\end{minipage}
\begin{minipage}{0.48\textwidth}
 Marginal change in the cost function for a marginal change in load:
 \begin{equation}
  LMP_3 = \frac{\partial \mathcal{L}}{\partial \xi_3} = -\nu -\lambda_{13}^+ PTDF_{13,3} \notag
 \end{equation}
\end{minipage}
\begin{center}
 To find $LMP_3$ I need $\nu$ and $\lambda_{13}^+$ \\
 How do I find $\nu$ and $\lambda_{13}^+$?
\end{center}

\begin{itemize}
 \item Solve the linear system with 2 equations and 2 unknowns: $\nu$ and $\lambda_{13}^+$
\end{itemize}
\begin{align}
   & c_2+\nu +\lambda_{13}^+ PTDF_{13,2} =0  \notag \\
   & c_3+\nu+\lambda_{13}^+ PTDF_{13,3} =0 \notag
\end{align}
\rule{\textwidth}{0.5pt}
\begin{itemize}
 \item What can we say about the LMPs on different buses?
\end{itemize}
\begin{equation}
 LMP_i =  -\nu -\lambda_{13}^+ PTDF_{13,i} \notag
\end{equation}
\begin{itemize}
 \item If there is a congestion, the LMPs are no longer the same on every bus. They are dependent on the congestion!
\end{itemize}


\bibliographystyle{IEEEtran}
\bibliography{diss_biblio}
\addcontentsline{toc}{chapter}{Bibliography}

\end{document}